\documentstyle[epsfig]{mn}

\title[The K--$z$ diagram for BCGs in X-ray clusters]
{The K band Hubble diagram for brightest cluster galaxies in X-ray
clusters}

\author[C.A. Collins and R.G. Mann]
{C.A. Collins$^1$ and R.G. Mann$^{2,3}$\\
$^1$Astrophysics Research Institute, School of Engineering, Liverpool John 
Moores University, Byrom Street, Liverpool L3 3AF\\
$^2$Astronomy Unit, 
Queen Mary and Westfield College, Mile End Road, London E1 4NS\\
$^3$Present Address: Astrophysics Group, Blackett Laboratory, 
Imperial College, Prince Consort Road, London SW7 2AZ}

\newcommand{\beq}{\begin{equation}}
\newcommand{\eeq}{\end{equation}}
\newcommand{\bfi}{\begin{figure}}
\newcommand{\efi}{\end{figure}}
\newcommand{\bit}{\begin{itemize}}
\newcommand{\eit}{\end{itemize}}
\newcommand{\rmd}{{\rm d}}
\newcommand{\myref}[1]{\noindent \hangindent=0.5in \hangafter=1 #1 \par}

\begin{document}

\maketitle
\begin{abstract}
This paper concerns the K band Hubble diagram for the brightest cluster
galaxies (BCGs) in a sample of X-ray clusters covering the redshift 
range $0.05<z<0.8$. We show that BCGs in clusters of high X-ray
luminosity are excellent standard candles: the intrinsic dispersion 
in the raw K band absolute magnitudes of BCGs in clusters with
$L_{\rm X} > 2.3 \times 10^{44}$ erg s$^{-1}$ (in the
0.3 - 3.5 keV band) is no more than 0.22 mag, and is not significantly reduced
by correcting for the BCG structure parameter, $\alpha$, or for
X-ray luminosity. This is the smallest scatter in the absolute
magnitudes of any single class of galaxy and demonstrates the
homogeneity of BCGs in high-$L_{\rm X}$ clusters. By contrast, we
find that the brightest members of low-$L_{\rm X}$ systems display
a wider dispersion ($\sim 0.5$ mag) in absolute magnitude than 
commonly seen in previous studies, which arises from the inclusion,
in X-ray flux-limited samples, of poor clusters and groups which are
usually omitted from low redshift studies of BCGs in optically rich
clusters.

Spectral synthesis models reveal the insensitivity of K band light to
galaxy evolution, and this insensitivity, 
coupled with the tightness of its Hubble relation, and the lack of 
evidence of significant growth by merging (shown by the absence of
a correlation between BCG structure parameter, $\alpha$, and redshift),
makes our sample of BCGs in high-$L_{\rm X}$ clusters  ideal
for estimating the cosmological parameters $\Omega_{\rm M}$ and 
$\Omega_{\Lambda}$, 
free from many of the problems that have bedevilled previous attempts using 
BCGs. The BCGs in our high-$L_{\rm X}$ clusters yield a
value of $\Omega_{\rm M}=0.28\pm0.24$ if the 
cosmological constant $\Lambda=0$. For a flat Universe we find 
$\Omega_{\rm M}=0.55^{+0.14}_{-0.15}$ with a 95 per cent confidence
 upper limit to the cosmological constant corresponding to
$\Omega_{\Lambda}<0.73$. These results are discussed in the
context of other methods used to constrain the density of the Universe,
such as Type Ia supernovae.

\end{abstract} 
\section{INTRODUCTION}

The Hubble (redshift--magnitude) diagram for brightest cluster galaxies
(BCGs) is a classic cosmological tool (recently reviewed by Sandage
1995). Sandage and collaborators (e.g. Sandage 1972a,b; 
Sandage \& Hardy 1973; Sandage, Kristian \& Westphal 1976)  used the 
Hubble diagram for BCGs to verify
Hubble's linear redshift--distance law out to a distance one hundred
times greater than that probed by the initial bright galaxy sample of 
Hubble (1929), and attempted to measure the deceleration parameter
from its deviation from the linear Hubble law.
More recently, the BCG redshift--magnitude relation has
appeared in studies of the formation and evolution of
elliptical galaxies (Arag\'{o}n--Salamanca et al. 1993) and of large-scale
bulk flows in the Universe (Lauer \& Postman 1994; Postman \& Lauer
1995; Hudson \& Ebeling 1997).

Underlying all this work is the fact that BCGs appear to be good 
standard candles after correction  for systematic effects and the dependence
of absolute magnitude on galaxy structure and environment, with their 
aperture luminosities at low redshift
showing a dispersion of only $\sim0.3$ mag (Sandage 1988). This small 
dispersion is  all the more remarkable since BCGs do
not appear to be a particularly homogeneous class of objects at first
sight: some
exhibit the large, extended envelopes that make them cD galaxies; many
(up to half according to Hoessel \& Schneider 1985) have multiple
nuclei; and their immediate local environments vary from relative
isolation in the cluster core to interaction and merging, in the case
of dumbbell galaxies. BCGs tend to dominate their host clusters
visually, but little is known of their place in the formation and
evolution of the clusters, or of their own origins. 

Answers to some
of these questions may result from extending the Hubble diagram for
BCGs to higher redshifts, but there are several problems with that.
Firstly, as one moves to higher redshifts, optical wavebands become
increasingly sensitive to the effects of star formation, and
K-corrections become large and uncertain. 
 This problem may be circumvented, as discussed in detail below, by 
working in the
K band, which continues principally to sample the mature stellar
population out to high redshifts: the K band Hubble diagram for a
small sample of BCGs was presented by Arag\'{o}n--Salamanca et
al. (1993). 

A more serious problem is that of cluster
selection. In the past, cluster samples have been selected from
optical surveys, on the basis of identifying enhancements in the
surface density of galaxies above a fluctuating background. Such
a procedure suffers from the well-known projection effects caused by chance 
alignments of galaxies
along the line-of-sight (van Haarlem, Frenk \& White
1997, and references therein). This contamination will be more severe at
fainter magnitudes, thus strongly discouraging the use of this method
for detecting high-redshift clusters. A solution to this problem is
provided by X-ray cluster selection. Most {\em bona fide\/} clusters contain
hot gas in their cores, emitting a copious flux of X-rays and making
them visible to large distances. The intensity of the thermal bremsstrahlung 
X-ray emission from  a
cluster is directly related to the depth of its gravitational potential
well, unlike its optical galaxy richness, while the compactness of the
X-ray-emitting
region, compared to the extent of the galaxy distribution, also means
that projection effects are minimal in comparison to those arising in
optical cluster selection (Romer et al. 1994; van Haarlem et al. 1997).

These concerns motivated the current study, the long-term goal of which
is to investigate the formation and evolution of BCGs in host clusters 
with known X-ray luminosities, by studying their physical properties over a 
wide span of cosmological time. In this paper we
concentrate on extending the infrared Hubble diagram for BCGs to higher
redshifts than studied before using K band images of a large sample of
BCGs in X-ray clusters with \mbox{$z \la 0.8$}. The plan of the remainder of 
this paper is as follows. In
Section 2 we briefly describe our cluster sample, the observations
made and their reduction, leading to the presentation of the 
uncorrected $K-z$ diagram for our sample in Section 3, 
followed by a discussion of K-correction models for our BCGs in Section 4.
We consider the physical
properties of the BCGs and their host clusters in Section 5 and look
for correlations between them in Section 6. These correlations are
used to produce a corrected Hubble diagram in Section 7, which is used
to derive constraints on the cosmological parameters $\Omega_{\rm M}$
and $\Omega_\Lambda$ in Section 8.
In Section 9 we discuss the results of this paper and  
present the
conclusions we draw from them. An Appendix tabulates some statistical
results omitted from the main body of the text.

\section{THE DATA}

\subsection{The X-ray Cluster Sample}

\begin{figure}

\epsfig{file=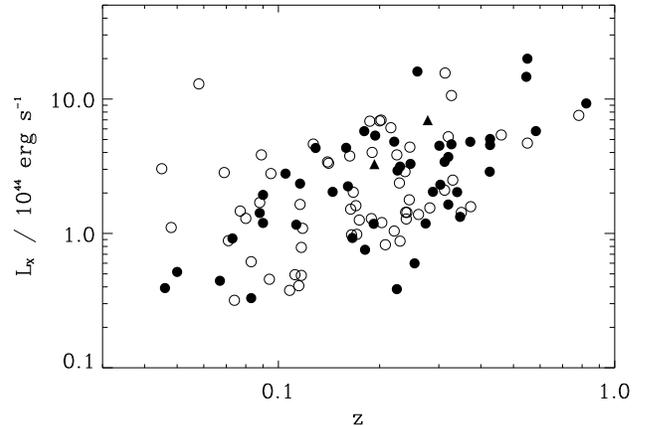,angle=0,width=9cm}

\caption{The luminosities of the EMSS clusters (in the 0.3--3.5 keV
passband) against redshift. The 
filled circles are the clusters imaged at K and discussed in this paper, 
while the empty circles are the rest of the EMSS cluster sample of
Gioia \& Luppino (1994). The two
triangles are the {\em ROSAT\/} clusters. Note that there 
is no single, sharp flux limit to the EMSS survey -- cluster
detections were based 
on pointed observations whose limiting sensitivity ranges from 
$5\times10^{-14}$ erg cm$^{-2}$ s$^{-1}$ to 
$3 \times 10^{-12}$ erg cm$^{-2}$ s$^{-1}$.}
 \label{emss_z_lx}

\end{figure}

The X-ray clusters for this work were chosen principally from the 
{\em Einstein\/} Medium Sensitivity Survey (Gioia et al. 1990, Stocke
et al. 1991, hereafter EMSS) catalogue of 104 X-ray selected clusters
(Gioia \& Luppino 1994), but supplemented by two clusters from the 
{\em ROSAT\/} All--Sky Survey (Voges 1992, Tr\"{u}mper 1993). 
Fig.~\ref{emss_z_lx} 
shows the X-ray luminosities listed in Gioia \& Luppino (1994) for all the 
EMSS clusters, as a function of redshift, compared with the sample  
studied here: the X-ray luminosities were calculated assuming that
$\Omega_{\rm M}=1$, $\Omega_\Lambda=0$ and the Hubble constant is
$H_0=50$ km s$^{-1}$ Mpc$^{-1}$. No attempt 
was made to observe a statistically complete subsample of EMSS clusters, 
although it is clear from Fig. \ref{emss_z_lx} that those
selected are broadly
representative of the range of X-ray luminosities for clusters within the 
parent survey: a two--sided Kolmogorov--Smirnov (K-S) test yields a
probability of 0.26 that the cumulative X-ray luminosity distributions
of the EMSS clusters imaged and not imaged would differ by more than
observed if both subsamples were drawn from the same parent
distribution. The corresponding K-S
test on the redshift distributions of EMSS cluster subsamples imaged and not
imaged yields a probability of only 0.04, reflecting the fact that
we preferentially
selected EMSS clusters at higher redshifts, where the luminosity
distance varies appreciably with cosmology and, thus,  constraints 
on  $\Omega_{\rm M}$ and $\Omega_\Lambda$ may be derived from the
BCG Hubble diagram.

\subsection{Observations}

Observations were made on 8-10 November 1994 and 20-22 April 1995
using the IRCAM3 infrared camera on the 3.8m United Kingdom Infrared 
Telescope (UKIRT): IRCAM3 contains a
$256\times256$ InSb array, with a pixel size of 0.286 arcsec. 
Two observing procedures were followed. For distant clusters, a series
of 100s (10 $\times$ 10s co-added) exposures were taken, with the 
galaxy offset in a five-point jitter pattern within the
initial field of view of the chip. For nearby clusters,
where the BCG covered too large an area of the chip to allow accurate
sky level determination by this method, sky exposures were obtained 
after each galaxy exposure, allowing the jitter pattern for the sky frames 
\mbox{(four-,$$} or eight-point) to be built-up simultaneously with
that of the 
jitter pattern for the galaxy frames \mbox{$$(five-,} or nine-point). 
Typical total
on-source integration times used were $\sim1000$s for $z\la 0.2$,
1500-3000s for $0.2 \la z \la 0.4$ and 4500s above $z=0.5$. Dark
frames were taken regularly throughout the nights, as were standard
star observations, using sources from the UKIRT Faint Standards list
(Casali \& Hawarden 1992). 

\subsection{Data Reduction}

The data were reduced using standard {\sc iraf}\footnote{{\sc iraf} is
distributed by National Optical Astronomy Observatories, which is
operated by the Association of Universities for Research in Astronomy,
Inc., under contract to the National Science Foundation.} routines and the
following procedure. Appropriate dark frames were subtracted from each
galaxy or sky image, and the resulting frames were multiplied by a
mask, flagging known bad pixels in the array to be ignored
subsequently, before normalising the frames to unit median: we found the
median to be a more stable measure of the sky level than the mode.
Flat-field frames were constructed, by median filtering the galaxy
frames (for distant clusters) or sky frames (for nearby clusters),
and these flat frames were normalised to unit median. The galaxy
frames were then divided by the appropriate flat field frame and
final mosaic images produced by cross-correlating the individual
flat-fielded galaxy 
frames to determine their relative offsets. A similar procedure was
used to reduce the standard star frames.

A total of 52 BCGs were observed. Two of these, (those in MS1401.9+0437 and
MS1520.1+3002), did not yield photometric quality data and were dropped
from our sample, as were two  others (MS1209.0+3917 and MS1333.3+1725)
which have dubious redshift identifications in Gioia \& Luppino (1994). 
A fifth BCG, in MS0440.5+0204, was
subsequently omitted, because it is close to a bright star, which
falls within our chosen photometric aperture. This left us with a sample
of 47 BCGs: note that Stocke et al. (1991) and Gioia \& Luppino (1994)
suggest that the two EMSS sources MS2215.7-0404 and MS2216.0-0401
may be part of the extended X-ray emission from a single cluster, but 
we have treated them as two distinct objects. We assume
cosmological matter and vacuum energy densities of $\Omega_{\rm M}=1$
and $\Omega_\Lambda=0$, respectively, and 
$h=0.5$ (where $H_0 = 100h\,{\rm km s}^{-1}{\rm Mpc}^{-1}$  
is the Hubble constant) throughout this paper, unless stated otherwise.

\begin{figure}

\epsfig{file=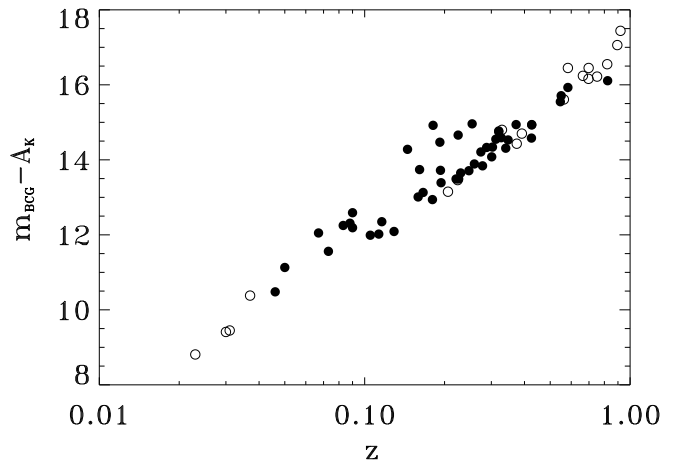,angle=0,width=9cm}

	\caption{The raw $K-z$ diagram. The filled circles show the apparent
K band magnitudes of the 47 BCGs in our X-ray sample, while the
empty circles are the BCGs from the optical sample of
Arag\'{o}n-Salamanca et al. (1993), excluding the one BCG in common
between the two samples. The magnitudes are corrected for
Galactic extinction, but have not been K-corrected.}
 \label{raw_k_z}

\end{figure}

\section{THE RAW $\bmath{K}$--$\bmath{\lowercase{z}}$ DIAGRAM}

We obtained K band aperture magnitudes for our BCGs using
the {\sc iraf} package {\sc apphot}, with a fixed metric 
aperture of diameter 50 kpc: in
the few cases, where there were two similarly bright galaxies in the
cluster, the BCG was taken to be the brighter of the two as judged
by this magnitude. Airmass corrections were made using the standard
median Mauna Kea K band extinction value of 0.088 mag/airmass.

The choice of aperture was made on the basis of two considerations.
Firstly, it allows direct comparison with the results of 
Arag\'{o}n-Salamanca et al. (1993) who studied BCGs in nineteen
optically-selected clusters. Secondly, as discussed in Section 5.1,
we require a sufficiently large metric aperture to allow accurate
estimation of the BCG structure parameter, $\alpha$, for our more
distant clusters, given the seeing conditions obtaining during our
runs.

From comparing frames of the same object taken at
different times on the same night, on different nights of the same run
and between the two runs, we obtain formal statistical errors of 0.04 to
0.07 mags on our 50 kpc magnitudes, except for 11 BCGs observed on one
particular night for which greater uncertainty in our photometric
calibration produced uncertainties of 0.1 mags. For the one object 
(MS0015.9+1609 $\equiv$ Cl0016+16) we have in common with
Arag\'{o}n-Salamanca et al. (1993) they measure an aperture magnitude
that agrees with ours to well within these estimated errors  (15.56
mag compared to our 15.58 mag). In addition to these statistical
errors, there are likely to be systematic errors resulting, for
example, from contamination from other cluster galaxies falling within
the photometric aperture: we make no attempt to estimate these
(necessarily very uncertain) corrections.
The aperture magnitude data set is given in Table 2 and
the resultant raw $K-z$ diagram is shown in Fig. \ref{raw_k_z}. 
It is clear from Fig. \ref{raw_k_z} that there is a large scatter in the raw
$K-z$ diagram for the BCGs in our X-ray cluster sample, and that
most of the scatter comes from the lower-$z$ half of the sample:
before expressing that scatter in terms of the rms dispersion in the
absolute magnitudes of the galaxies, and investigating its origin, we
must discuss how to K--correct the galaxy magnitudes, which we do in
the next Section.

\section{Interpreting the BCG K band magnitudes}

There is a consensus developing that cluster ellipticals are old, for
example: Charlot \& Silk (1994) model the evolution of spectral
indicators of E/S0 galaxies in low-redshift ($z<0.4$) clusters, and
suggest that a couple of per cent at most of their stars have been
made in the past 2.5 Gyr;
Bender, Ziegler \& Bruzual (1996) use Mg$_b-\sigma$ data for 
16 cluster ellipticals in three clusters at $z\simeq0.37$ to show that
the majority of their stars must have formed at $z>2$, and that the most
luminous galaxies may have formed at $z>4$;
and Ellis et al. (1997) use rest--frame UV--optical photometry
of ellipticals in $z\sim0.5$ clusters obtained with the {\em Hubble
Space Telescope\/} to show that the bulk of the star formation in the 
ellipticals in dense clusters was completed before $z\sim3$ in
conventional cosmologies. If we assume that the same goes for our BCGs
as for cluster ellipticals in general, then 
our photometric data are easy to interpret, since, by choosing the
K band, we are primarily sampling the passively evolving mature
stellar population over our full redshift range, and the resultant
K--correction will be insensitive to the exact age of the galaxy:
we take the K--correction, $K(z)$, as encompassing the bandwidth, 
band--shifting and evolution terms (e.g. Sandage 1995), defining it so
that the absolute magnitude, $M$, of a galaxy observed to have an
apparent magnitude $m$ at redshift $z$ is given by
\mbox{$M=m+5-5\log_{10}(d_{\rm L})-K(z)$}, where $d_{\rm L}(z,\Omega_{\rm M},
\Omega_\Lambda, H_0)$ is the luminosity distance (in pc), and where we 
have neglected the effect of Galactic extinction.

In the light of these results, we take as our
default K--correction model that resulting from a Bruzual \&
Charlot GISSEL (Bruzual \& Charlot 1993) model (1995 version: see  
Charlot, Worthey \& Bressan
1996) in which our BCGs form in a 1 Gyr burst (with a Salpeter 1955
IMF, over the range \mbox{$0.1 \leq M/M_\odot \leq 125$}) at a redshift
$z_{\rm f}=3$, in a Universe with $\Omega_{\rm M}=1$, $\Omega_\Lambda=0$,
and $h=0.5$. This can be approximated to better than 0.02 mag over
the range \mbox{$0 \leq z \leq 1$} by a fifth-order polynomial in $z$, 
given by
\begin{equation}
K(z)=-2.476z-0.5809z^2+14.04z^3-21.10z^4+9.147z^5.
\label{kzmodel}
\end{equation}
This is computed by approximating the K band filter by a top-hat
between 2.0 and 2.45 $\mu$m: the computed K-correction changes by
$\la0.02$ mag over $0 \leq z \leq 1$ if instead we fit the 
UKIRT K band filter (S.K. Leggett, private communication)
with a ninth-order polynomial, which accurately follows variations in
its transmission from 1.9 to 2.5 $\mu$m.
The model of equation (1) agrees to better than 0.05 mag with the mean
empirical K band
K--correction derived by Bershady (1995) from the large sample of
$z\la 0.3$ field galaxies constructed by Bershady et al. (1994), which
is very uniform, having an rms dispersion of $\sim0.05$ mag over the full
range of field galaxy classes.

\begin{figure}

\epsfig{file=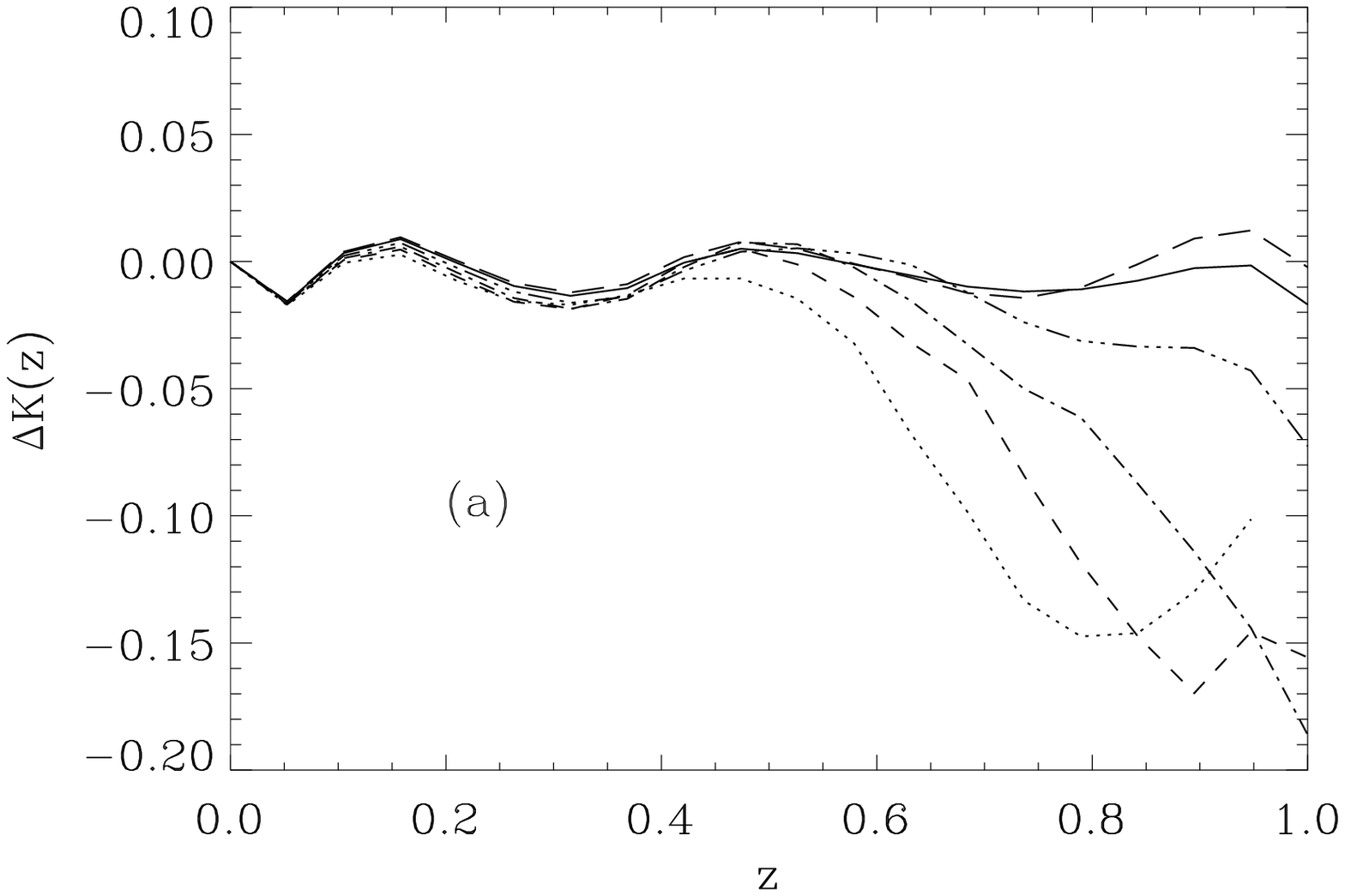,angle=0,width=9cm}

\epsfig{file=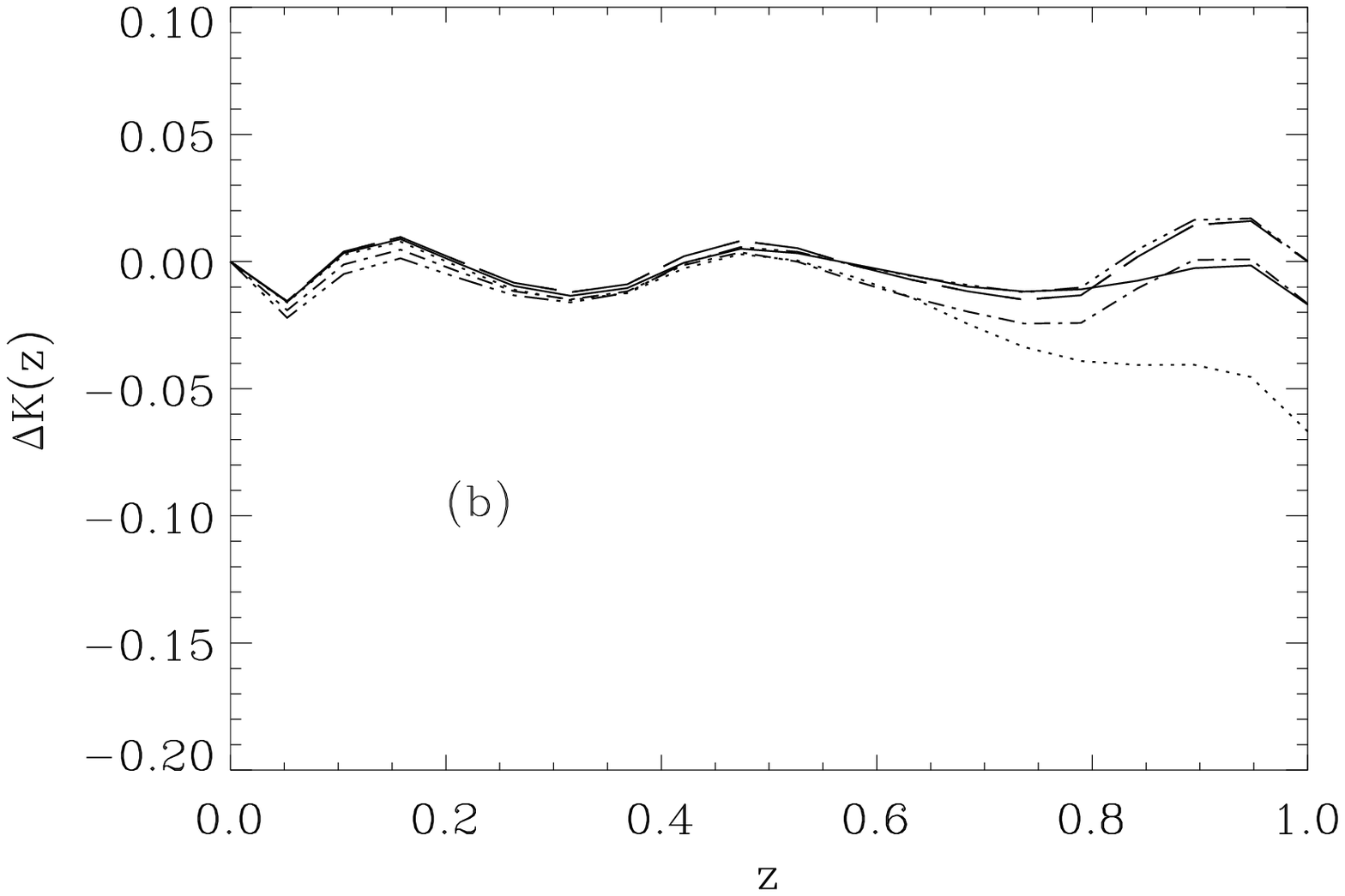,angle=0,width=9cm}

\caption{The difference between the default K band K--correction model of
equation (1) and other models, generated using Bruzual \&
Charlot's 1995 GISSEL library (see Charlot et al. 1996), all
using a Salpeter (1955) IMF and mass range  $0.1 \leq M/M_\odot
\leq 125$ unless otherwise stated: the oscillating form of these
$\Delta K(z)$ curves is caused by the use of Chebyshev polynomials
to produce the fit given in equation (1).
In (a) we consider bursts of star formation of 1 Gyr duration
at $z_{\rm f}=$ 1 (dotted line), 1.25 (dashed line), 1.5
(dot-dashed line), 2 (dot-dot-dot-dashed line), 3 (solid lines)
and  4 (long-dashed lines): the $K(z)$ curve does not change
significantly for $z_{\rm f} > 4$. In  (b) we show bursts at $z_{\rm f}=3$
with: (i) 1 Gyr duration (solid line, as in (a)); (ii) exponential
decaying star formation rate with $\tau=1$ Gyr (dotted line); (iii)
instantaneous burst (dashed line); (iv) instantaneous burst with
mass range  $0.1 \leq M/M_\odot \leq 2.5$ (long-dashed line); 
(v) instantaneous
burst with Scalo (1986) IMF (dot-dashed line); and (vi) instantaneous
burst with Miller \& Scalo (1979) IMF (dot-dot-dot-dashed line). We
assume $\Omega_{\rm M}=1,\; \Omega_\Lambda=0$ and $h=0.5$ in all
cases.The pronounced dips in $K(z)$ at $z$=0.7--1.0 for $z_{\rm f}$=
1.0--1.5 are caused by AGB stars: their appearance is short-lived, and
pushed to $z>1$ if $z_{\rm f} \geq 1.5$. We only have one BCG at
$z>0.6$, so their effect on our results will be negligible.}

\label{kofz_models}

\end{figure}

To illustrate the insensitivity of  K band K--corrections to 
star formation history we show in Fig.~\ref{kofz_models} the 
difference between the default model of equation (1)
and the $K(z)$
curves resulting from (a) varying $z_{\rm f}$, and (b) varying the IMF and
form of the burst of star formation. 
It is clear that the age of the stellar
population is the most important factor in determining the
K--correction, but the variation of the K--correction with age only 
becomes noticable above $z\sim0.4$, and is pushed to higher redshifts
as $z_{\rm f}$ increases: as long as the bulk of a galaxy's stars were
formed before $z_{\rm f}=1.0$, then its K--correction is known to 0.05 or
better if its redshift is  $z\leq 0.6$ (we only have one BCG more
distant than that) and known to that accuracy out to $z\sim0.8$ if $z_{\rm f} >
1.5$. Fig.~\ref{kofz_models} (b) shows that $K(z)$ is very insensitive
to the IMF and duration of the burst of star formation.

If we convert the apparent magnitudes in Fig.~\ref{raw_k_z} to absolute
magnitudes, $M_{\rm K}$, using the K-correction model of  
equation (1) then the K-corrected mean
absolute magnitude of the full 47 galaxy sample  is
\mbox{$\bar{M}_{\rm K}=-26.40$ mag}, and the rms dispersion about
that mean is $\sigma=0.47$ mag. If we consider the higher-redshift half
of the sample, (i.e. the 23 galaxies with $z>0.227$) then these figures
become \mbox{$\bar{M}_{\rm K}=-26.53$ mag} and $\sigma=0.30$: this
reduction in the rms dispersion in $\bar{M}_{\rm K}$ is significant at
the 98.6 per cent level, since only  1399 out of 10$^5$ 23-galaxy random
subsamples of the full BCG sample give $\sigma$ values less than 0.30. 
Rather than assuming that the BCGs were formed at a single redshift, 
$z_{\rm f}$, we could follow Glazebrook et
al. (1995) and assume that the stellar populations of our BCGs are of
a constant age,
irrespective of the redshift at which they are observed: that
procedure leads to very similar results as those of the K--correction of
equation (1)  as the age of the BCGs is varied over the range
$\sim1-10$ Gyr (Glazebrook et al. 1995). The constant $z_{\rm f}$ and 
constant age assumptions both represent crude treatments of the star
formation history of our BCGs, but the essential point here is that
the very weak dependence of $K(z)$ to the exact formation redshift
or exact age  illustrates how insensitive near-infrared K--corrections
are to the details of the star formation history of galaxies, and, hence,
show the power of K band observations of early-type galaxies with
a narrow intrinsic spread in absolute magnitude to act as standard candles.

It might be objected that the passive evolution models discussed
in this Section are not appropriate for BCGs, since they might have
experienced more recent star formation, either due to cooling flow
activity or merging with other cluster galaxies. Allen (1995)
studied six cooling flow clusters, finding large UV/blue continua and
strong emission-lines, such as ${\rm H}\alpha$ and $[{\rm OII}]$, which can be
interpreted as emission from O stars. In order to test for ongoing
star formation we have examined the difference in absolute magnitudes
between the EMSS
BCGs in our sample which show some evidence for star-forming activity in
their nucleus and the rest. From the EMSS cluster catalogue (Gioia \&
Luppino 1994) the BCGs in MS0302.7, MS0419.0, MS0537.1, MS0839.8,
MS0955.7, MS1004.2, MS1125.3, MS1224.7 and MS1455.0 all have
$[{\rm OII}]$, while MS0015.9 has many galaxies with  ``E+A'' type spectra.
Despite this, all ten lie very close to the mean Hubble relation shown
in Fig.~\ref{raw_k_z}. A two-sided K-S test for BCGs in our sample
with and without strong emission lines yields a probability of
0.17 that the difference in absolute magnitudes between the two
populations would arise if the two subsamples are drawn from the same
parent population. This result suggests that the effect of any ongoing
star formation in the centres of BCGs on the K band magnitudes integrated
over the 50 kpc diameter aperture  is small. Such an interpretation
of the spatial extent of any star formation in these systems is consistent
with a number of studies which have concluded that optical
colour gradients, 
indicative of active star formation, are either completely absent 
in BCGs (Andreon et al. 1992)  or are confined to their inner regions 
(McNamara \& O'Connell 1992). 
 
Some BCGs are radio galaxies, so another possible concern is that some
of the K band light from our BCGs comes from AGN, rather than from
stars (Eales et al. 1997, and references therein). Gioia \& Luppino
(1994) list radio detections for only ten of the BCGs in our sample,
and only six of these are in the high X-ray luminosity clusters in
which we shall be particularly interested: a K-S test for the absolute
magnitude distributions of our BCGs in high X-ray luminosity clusters
yields a probability of 0.78 that BCGs with and without radio
detections are drawn from the same parent population.

To summarise, there is no evidence to suggest we should not interpret
the K band photometry of our BCGs as arising from the  passive
evolution of a mature stellar population formed at $z \ga3$ similar to
that advocated by Ellis et al. (1997) and others for cluster ellipticals in
general. We shall, therefore, adopt equation (1) as our
preferred K--correction, but will return, in Section 8, to discuss the
effect that varying this assumed $K(z)$ relation has on the
constraints on $\Omega_{\rm M}$ and $\Omega_\Lambda$ we deduce from
the BCG Hubble diagfram.

\section{PROPERTIES OF BCGS AND THEIR HOST CLUSTERS}

The scatter in the raw Hubble diagram shown in Fig. \ref{raw_k_z} should 
come as no surprise, since it is well known that there exist strong
correlations between the optical luminosities of BCGs and properties 
both of themselves, and of their host clusters, and that it is only
after correction for these that BCGs are revealed to be
good standard candles. In this Section we investigate some of the physical
properties of the BCGs and their host clusters likely to be correlated
with BCG luminosity.

\subsection{The BCG structure parameter}

Sandage (1972a,b) found that the luminosities of BCGs correlate with
the richnesses and Bautz-Morgan (Bautz \& Morgan 1970) type of their
host clusters, and Hoessel (1980) showed
how this could be expressed as a correlation between
$L_{\rm m}$, the BCG luminosity within a metric aperture of
radius $r_{\rm m}$, and \mbox{$\alpha \equiv [ \rmd \log (L_{\rm m}) / 
\rmd \log (r)] |_{r=r_{\rm m}}$}, the logarithmic slope of the
BCG growth curve. Correction for the correlation with the structure
parameter, $\alpha$, greatly reduces the dispersion in BCG absolute
magnitudes, and the $L_{\rm m}-\alpha$ relation forms the basis of
the recent use of BCGs as streaming velocity probes (Lauer \& Postman 1994;
Postman \& Lauer 1995; Hudson \& Ebeling 1997).

To measure the structure parameters of our BCGs, we computed the growth
curve for each BCG from luminosity measurements in a
set of apertures with radii increasing in steps of 1 pixel out to 100
pixels or 30 arcsec. Each curve was then modelled using a Hermite polynomial, 
which was fitted over the full range of the growth curve. The value of 
$\alpha$ was then determined for each BCG by evaluating the logarithmic 
derivative of the polynomial fit to its growth curve at an angle
$\theta_{\rm m}$, which corresponds to a radius $r_{\rm m}=25$ kpc, and is 
given by
\beq
\theta_{\rm m}(\rm{arcsec})=0.43 \frac{(1+z)^2}{[1-(1+z)^{-1/2}]},
\label{theta_m}
\eeq
for our assumed cosmology, with $\Omega_{\rm M}=1$, $\Omega_\Lambda=0$
and $h=0.5$. There are three principal 
sources of potential systematic error in our $\alpha$ estimates:

\begin{enumerate}

\item Incorrect background subtraction. In all cases, our galaxies are
small enough that an adequate portion of
uncontaminated image was available to estimate the sky level. We estimated 
what effect residual variations in the sky background have on 
the estimated structure parameter  by recalculating the sky level using more
than one uncontaminated part of each image.  These tests were carried out 
on a random subset of 28 BCGs in our sample. The results indicated that  
uncertainties in sky subraction give rise to typical 1$\sigma$ errors of
about 0.06 which is $\sim10$ per cent on average: this
dominates the uncertainty in the estimation of  the structure 
parameter. 

\item The presence of contaminating sources within the metric
aperture. For 7 sources a star or second galaxy was found to lie
within the circle corresponding to a projected 
distance of 25 kpc from the centre of the BCG. In these cases the 
contaminating 
sources were masked out, and the values of the pixels in the masked region, 
(which, in all cases, were close to 25 kpc away from the BCG), were 
replaced with the
value of the pixel located at the same distance from the BCG but diametrically 
opposite the contaminating source. In tests with non-contaminated
images, this process introduced very little additional error.

\begin{table}


\caption{Seeing disk convolution results: the mean increase in
$\alpha$ for two values of $r_{\rm m}$, as a function of redshift.}

\begin{tabular}{lccc}
$r_{\rm m}$ & mean $\Delta \alpha$ &mean $\Delta \alpha$ & mean $\Delta \alpha$\\
(kpc) & $0.2<z \leq 0.3$ & $0.3 <z \leq 0.4$   & $0.4 < z <0.8$ \\ \hline
8    & 0.04  & 0.05  &  0.10 \\
25  & 0.02 & 0.01  &  0.03 \\
\end{tabular}

\end{table}

\item Profile broadening due to seeing. In the inner region of the BCG
growth curves, the broadening effect
of seeing causes the $\alpha$ values to be systematically overestimated.
Over the course of our observations the seeing varied between 0.9 and 1.4
arcsec, with a median seeing of 1.1 arcsec: a seeing disk of this size
is equivalent to a physical size of 5 kpc at $z=0.2$ and 8 kpc at $z=0.5$. To
investigate the size of this effect we smoothed our images by convolving
them with a Gaussian filter with  a FWHM of 1.1 arcsec and
re-measured the structure parameters. The mean increase in the measured
$\alpha$ parameter computed at \mbox{$r_{\rm m}=8 h^{-1}$ kpc} and
\mbox{$r_{\rm m}=25 h^{-1}$ kpc} in  redshift bins is shown in Table 1.
It is clear from Table 1 that the estimated corrections to 
the $\alpha$ parameter
measured in a 50 kpc diameter aperture are small compared to other 
uncertainties, 
and there is little variation with redshift, so it is not necessary to 
correct our data for the effects of seeing. The advantages of using a
larger aperture can also be seen, since the systematic corrections are 
significantly smaller for a 50 kpc aperture compared with those for a 16 kpc 
aperture. The adopted $\alpha$ values for our BCGs are tabulated in Table 2. 

\end{enumerate}

\subsection{Host cluster X-ray luminosities}

Hudson \& Ebeling (1997) have recently argued that the residuals in
the $L_{\rm
m}-\alpha$ relation correlate with the X-ray luminosity of the host
cluster, so that the $L_{\rm m}-\alpha$ relation does not fully remove
the environmental dependence on BCG luminosity: the potential use of
the cluster X-ray luminosity and/or temperature in reducing the
scatter in the BCG Hubble diagram was also discussed by Edge (1991).
In Table 2 we list the
X-ray luminosities of the host clusters of our BCG sample.
The EMSS  X-ray luminosities for the 45 EMSS clusters are taken from Gioia
\& Luppino (1994), and are quoted for the energy pass band of the
Imaging Proportional Counter ($0.3-3.5$ keV). 
We supplement that information with the
luminosities for two {\em ROSAT\/} clusters
(R84155 and R843053) which were discovered as part of an investigation of 
large-scale structure in the southern hemisphere  
(see Romer et al. 1994). The X-ray
luminosities of these additional clusters have been corrected for 
extinction by using H{\sc i} column densities interpolated from 
Stark et al. (1992), then 
K-corrected and transformed from the 
{\em ROSAT\/} Position Sensitive
Proportional Counter energy pass band 
($0.1-2.4$ keV) to the EMSS energy range assuming a 6 keV thermal 
bremsstrahlung spectrum: the pass band correction is of the form 
$L_{\rm X}(0.3-3.5 \; {\rm keV})=1.08 \times L_{\rm X}(0.1-2.4 \; {\rm keV})$.

\begin{table*}

\begin{minipage}{140mm}

\caption{Data for the BCGs in the X-ray cluster sample}

\begin{tabular}{lccccccccc}
Cluster & RA & dec & z & $L_{\rm X}$ & $m_{\rm BCG}$  & $A_K$ &$\alpha$ & 
$N$& $\hat{N}$  \\
\hline \\
 
MS0007.2-3532 & 00$^h$07$^m$14\fs5 & -35\degr33\arcmin11\farcs8 &
0.050 & 0.517 & 11.14 & 0.01 & 0.49 & 0 & -0.50 \\
MS0015.9+1609 & 00$^h$15$^m$58\fs3 & +16\degr09\arcmin34\farcs0 & 
0.546&14.639 & 15.58 & 0.03 & 0.57 & 5 & 3.15 \\
MS0301.7+1516 & 03$^h$01$^m$43\fs3 & +15\degr15\arcmin50\farcs6 &
0.083 &0.330 & 12.33 & 0.08 & 0.43 & 0  & -1.00\\
MS0302.5+1717 & 03$^h$02$^m$29\fs4 & +17\degr16\arcmin47\farcs6 &
 0.425&2.879 & 14.65 & 0.07 & 0.75 & 5 & 3.89 \\
MS0302.7+1658 & 03$^h$02$^m$43\fs2 & +16\degr58\arcmin27\farcs0 &
 0.426&5.043 & 15.01 & 0.07 & 0.67 & 4 & 2.51 \\
MS0419.0-3848 & 04$^h$18$^m$59\fs6 & -38\degr49\arcmin01\farcs2 &
 0.225&0.385 & 14.67 & 0.01 & 0.32 & 0  & -2.36 \\
MS0433.9+0957 & 04$^h$33$^m$58\fs4 & +09\degr57\arcmin36\farcs7 &
 0.159&4.335 & 13.11 & 0.10 & 0.67 & 4 & 3.14 \\
MS0451.6-0305 & 04$^h$51$^m$40\fs5 & -03\degr05\arcmin46\farcs0 &
 0.55&19.976 & 15.74 & 0.03 & 0.61 & 6 & 3.94\\
MS0537.1-2834 & 05$^h$37$^m$06\fs8 & -28\degr34\arcmin40\farcs6 &
 0.254&0.599 & 14.97 & 0.01 & 0.28 & 3 & 0.43\\
MS0821.5+0337 & 08$^h$21$^m$33\fs7 & +03\degr37\arcmin30\farcs3 &
 0.347&1.328 & 14.55 & 0.02 & 0.76 & 4 & 2.74\\
MS0839.8+2938 & 08$^h$39$^m$53\fs3 & +29\degr38\arcmin16\farcs0 &
 0.194&5.348 & 13.42 & 0.03 & 0.65 & 7 & 6.10\\
MS0849.7-0521 & 08$^h$49$^m$46\fs3 & -05\degr21\arcmin36\farcs5 &
 0.192&1.179 & 14.49 & 0.02 & 0.34 & 4 & 1.47\\
MS0904.5+1651 & 09$^h$04$^m$33\fs0 & +16\degr51\arcmin15\farcs0 &
 0.073&0.918 & 11.59 & 0.03 & 0.46 & 6  & 5.53\\
MS0906.5+1110 & 09$^h$06$^m$30\fs1 & +11\degr10\arcmin39\farcs3 &
 0.180&5.769 & 12.97 & 0.03 & 0.69 & 7 & 6.39 \\
MS0955.7-2635 & 09$^h$55$^m$45\fs2 & -26\degr35\arcmin56\farcs2 &
 0.145&2.039 & 14.32 & 0.04 & 0.13 & 6 & 2.68\\
MS1004.2+1238 & 10$^h$04$^m$12\fs7 & +12\degr38\arcmin15\farcs9 &
 0.166&0.925 & 13.16 & 0.03 & 0.81 & 4 & 3.14\\
MS1006.0+1202 & 10$^h$06$^m$07\fs3 & +12\degr02\arcmin20\farcs4 &
 0.221&4.819 & 13.52 & 0.03 & 0.74 & 9 & 8.17\\
MS1008.1-1224 & 10$^h$08$^m$05\fs4 & -12\degr25\arcmin07\farcs4 &
 0.301&4.493 & 14.12 & 0.04 & 0.73 & 8 & 7.17 \\
MS1054.4-0321 & 10$^h$54$^m$26\fs9 & -03\degr21\arcmin25\farcs3 &
 0.823&9.281 & 16.14 & 0.03 & 0.70 & 8 & 5.85 \\
MS1111.8-3754 & 11$^h$11$^m$49\fs8 & -37\degr54\arcmin55\farcs8 &
 0.129&4.325 & 12.16 & 0.07 & 0.70 & 2 & 1.61  \\
MS1125.3+4324 & 11$^h$25$^m$17\fs9 & +43\degr24\arcmin09\farcs5 &
 0.181&0.756 & 14.94 & 0.02 & 0.23 & 3 & -1.00 \\
MS1127.7-1418 & 11$^h$27$^m$52\fs3 & -14\degr18\arcmin19\farcs2 &
 0.105&2.786 & 12.02 & 0.03 & 0.90 & 6 & 5.55\\
MS1147.3+1103 & 11$^h$47$^m$18\fs2 & +11\degr03\arcmin15\farcs9 &
 0.303&2.304 & 14.36 & 0.02 & 0.61 & 3 & 1.76 \\
MS1208.7+3928 & 12$^h$08$^m$44\fs0 & +39\degr28\arcmin19\farcs0 &
0.340&2.030 & 14.32 & 0.01 & 0.55 & 3 & 1.95 \\
MS1224.7+2007 & 12$^h$24$^m$42\fs6 & +20\degr07\arcmin30\farcs0 &
 0.327&4.606 & 14.61 & 0.02 & 0.62 & 1 & -0.41 \\
MS1241.5+1710 & 12$^h$41$^m$31\fs6 & +17\degr10\arcmin06\farcs7 &
 0.312&3.411 & 14.56 & 0.01 & 0.44 & 5 & 3.57 \\
MS1253.9+0456 & 12$^h$53$^m$54\fs1 & +04\degr56\arcmin25\farcs7 &
 0.230&3.143 & 13.66 & 0.01 & 0.56 & 4 & 3.09 \\
MS1426.4+0158 & 14$^h$26$^m$26\fs7 & +01\degr58\arcmin36\farcs9 &
 0.320&3.707 & 14.77 & 0.02 & 0.58 & 4 & 2.35 \\
MS1455.0+2232 & 14$^h$55$^m$00\fs5 & +22\degr32\arcmin34\farcs7 &
 0.259&16.029  & 13.91 & 0.02 & 0.56 & 3 & 2.01 \\
MS1512.4+3647 & 15$^h$12$^m$25\fs9 & +36\degr47\arcmin26\farcs7 &
 0.372&4.807 & 14.95 & 0.01 & 0.82 & 8 & 6.38 \\
MS1522.0+3003 & 15$^h$22$^m$03\fs6 & +30\degr03\arcmin51\farcs1 &
 0.116&2.347 & 12.36 & 0.01 & 0.53 & 7 & 6.41 \\
MS1531.2+3118 & 15$^h$31$^m$14\fs1 & +31\degr18\arcmin42\farcs2 &
 0.067&0.444 & 12.07 & 0.02 & 0.35 & 5 & 3.95 \\
MS1532.5+0130 & 15$^h$32$^m$29\fs4 & +01\degr30\arcmin46\farcs3 &
 0.320&1.641 & 14.80 & 0.03 & 0.36 & 3 & 1.31 \\
MS1546.8+1132 & 15$^h$46$^m$52\fs0 & +11\degr32\arcmin25\farcs6 &
 0.226&2.937 & 13.52 & 0.03 & 0.52 & 4 & 3.19 \\
MS1558.5+3321 & 15$^h$58$^m$26\fs9 & +33\degr21\arcmin40\farcs7 &
 0.088&1.420 & 12.33 & 0.02 & 0.77 & 7 & 6.10 \\
MS1617.1+3237 & 16$^h$17$^m$08\fs8 & +32\degr37\arcmin52\farcs6 &
 0.274&1.185 & 14.22 & 0.01 & 0.79 & 6 & 4.76 \\
MS1618.9+2552 & 16$^h$18$^m$56\fs7 & +25\degr53\arcmin22\farcs3 &
 0.161&2.241 & 13.77 & 0.03 & 0.94 & 7 & 5.31 \\
MS1621.5+2640 & 16$^h$21$^m$32\fs2 & +26\degr41\arcmin06\farcs4 &
 0.426& 4.546 & 14.96 & 0.03 & 0.32 & 5 & 3.57 \\
MS2053.7-0449 & 20$^h$53$^m$44\fs0 & -04\degr49\arcmin24\farcs7 &
 0.583&5.775 & 15.96 & 0.03 & 0.51 & 4 & 1.70 \\
MS2124.7-2206 & 21$^h$24$^m$39\fs4 & -22\degr07\arcmin15\farcs2 &
 0.113&1.161 & 12.05 & 0.03 & 0.57 & 4 & 3.58 \\
MS2215.7-0404 & 22$^h$15$^m$41\fs3 & -04\degr04\arcmin25\farcs2 &
 0.090&1.196 & 12.63 & 0.04 & 0.60 & 4 & 2.75 \\ 
MS2216.0-0401 & 22$^h$16$^m$04\fs7 & -04\degr01\arcmin51\farcs9 &
 0.090&1.935 & 12.23 & 0.04 & 0.93 & 5 & 4.23 \\
MS2255.7+2039 & 22$^h$55$^m$40\fs7 & +20\degr39\arcmin04\farcs2 &
 0.288& 2.041 & 14.37 & 0.04 & 0.62 & 8 & 6.67 \\
MS2301.3+1506 & 23$^h$01$^m$17\fs1 & +15\degr06\arcmin49\farcs8 &
 0.247&3.291 & 13.75 & 0.04 & 0.52 & 3 & 2.10 \\
MS2354.4-3502 & 23$^h$54$^m$25\fs9 & -35\degr02\arcmin15\farcs8 &
 0.046&0.392 & 10.49 & 0.01 & 0.51 & 1 & 0.78 \\
R84155  &01$^h$42$^m$20\fs2 &-22\degr28\arcmin39\farcs8 & 0.278 &
 6.97 &13.85 & 0.01 & 0.55 & 5 & 4.14 \\
R843053  &23$^h$47$^m$47\fs7 & -24\degr52\arcmin30\farcs9 & 0.193
 & 3.28& 13.73 & 0.01 & 0.42 & 0 & -1.24 \\

\end{tabular}

\medskip
The names, positions, redshifts and X-ray luminosities of the EMSS
clusters (with prefix MS) are taken from Gioia \& Luppino (1994): the
RAs and decs. are given in B1950 coordinates and refer to the
brightest cluster member in the $V$ band; and the X-ray
luminosities, $L_{\rm X}$, are in units of 10$^{44}$ erg s$^{-1}$. $m_{\rm
BCG}$ is the BCG K band magnitude within a 50 kpc diameter aperture,
$A_K$ is the  K band extinction correction (computed using the
$N_{\rm H{\sc i}}$ maps of Stark et al. 1992, $R_{\rm V}=3.5$ and the mean
extinction law of Mathis 1990), $\alpha$ is the structure
parameter at 50 kpc diameter, while $N$ and
$\hat{N}$ are, respectively, the raw and corrected BCG near neighbour 
numbers. Clusters commonly known by another 
name are as follows: MS0015.9+1609 (Cl0016+16); MS0839.8+2938 (Zwicky 1883);
MS0904.5+1651 (Abell 744); MS0906.5+1110 (Abell 750); MS1006.0+1202
(Zwicky 2933); MS1127.7-1418 (Abell 1285); MS1253.9+0456 (Zwicky 5587);
MS1522.0+3003 (Abell 2069); MS1531.2+3118 (Abell 2092); MS1558.5+3321 
(Abell 2145); MS1618.9+2552 (Abell 2177); MS2255.7+2039 (Zwicky 8795); 
MS2301.3+1506 (Zwicky 8822); MS2354.4-3502 (Abell 4059); and R84155 (Abell 
2938)

\end{minipage}

\end{table*}

\subsection{Near neighbour analysis}

To test how well the correlations with $\alpha$ and $L_{\rm X}$ can 
express the environmental dependences of the BCG absolute magnitude,
we wanted also to study residual dependences on the richness of
the host cluster. Since our principal goal was to measure BCG magnitudes,
we did not image more of the host clusters than fell into the field of
view of the chip while imaging the BCG itself. In the most extreme
case, this means we can only count near neighbours out to a distance
of 100 kpc from the BCG:
this means that we are looking at the central core of the cluster,
rather than determining the richness of the cluster as a whole, for
which one would want to count neighbours to some appreciable fraction of
an Abell (1958) radius.

We used the {\sc starlink pisa} package (Draper \& Eaton 1996)
to detect near neighbours within a
projected distance of 100 kpc of the BCG centre, by searching for 
sets of at least ten contiguous pixels lying at least 2$\sigma$ above
the sky level and used the {\sc iraf apphot} package to estimate their
magnitudes: we removed `obvious' bright stars and foreground
galaxies by eye, although few BCGs had one of these falling within the
counting circle. From the sky noise level in each frame we were able
to deduce its limiting isophote, which we converted into a limiting
magnitude. We then computed, for each frame, the magnitude
difference between the 50 kpc BCG magnitude and the limiting magnitude
for that frame, finding that all frames went at least 4.66 magnitudes
deeper than the BCG magnitude. Our raw near neighbour count for each
BCG, $N$, was then taken to be the number of objects lying within a
projected distance of 100 kpc of the BCG and with a magnitude less
than 4.5 magnitudes fainter than the 50 kpc aperture magnitude of the
BCG. 

\begin{figure}

\epsfig{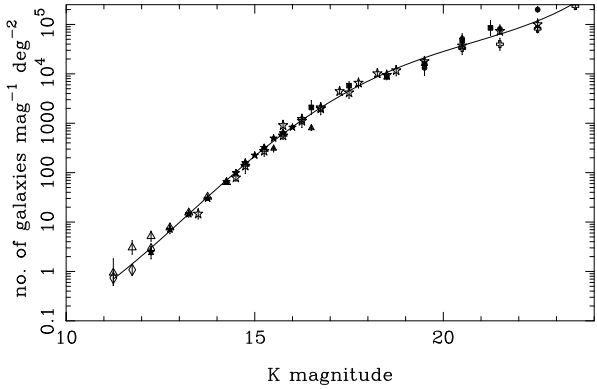}

\caption{K band field galaxy count data. The solid line
plots the fourth-order polynomial fit made to the count data, and
used to predict the background count for each BCG. The symbols for
the galaxy counts are as follows: diamonds, Mobasher et al. (1986); 
empty stars, Gardner et al. (1993); filled triangles, Glazebrook et
al. (1994); crosses, Djorgovski et al. (1995); squares, McLeod et al.
(1995); circles, Metcalfe et al. (1996); filled stars, Huang et al.
(1997); and empty triangles, Gardner et al. (1997).}

\label{kcounts}

\end{figure}

\begin{figure}

\epsfig{file=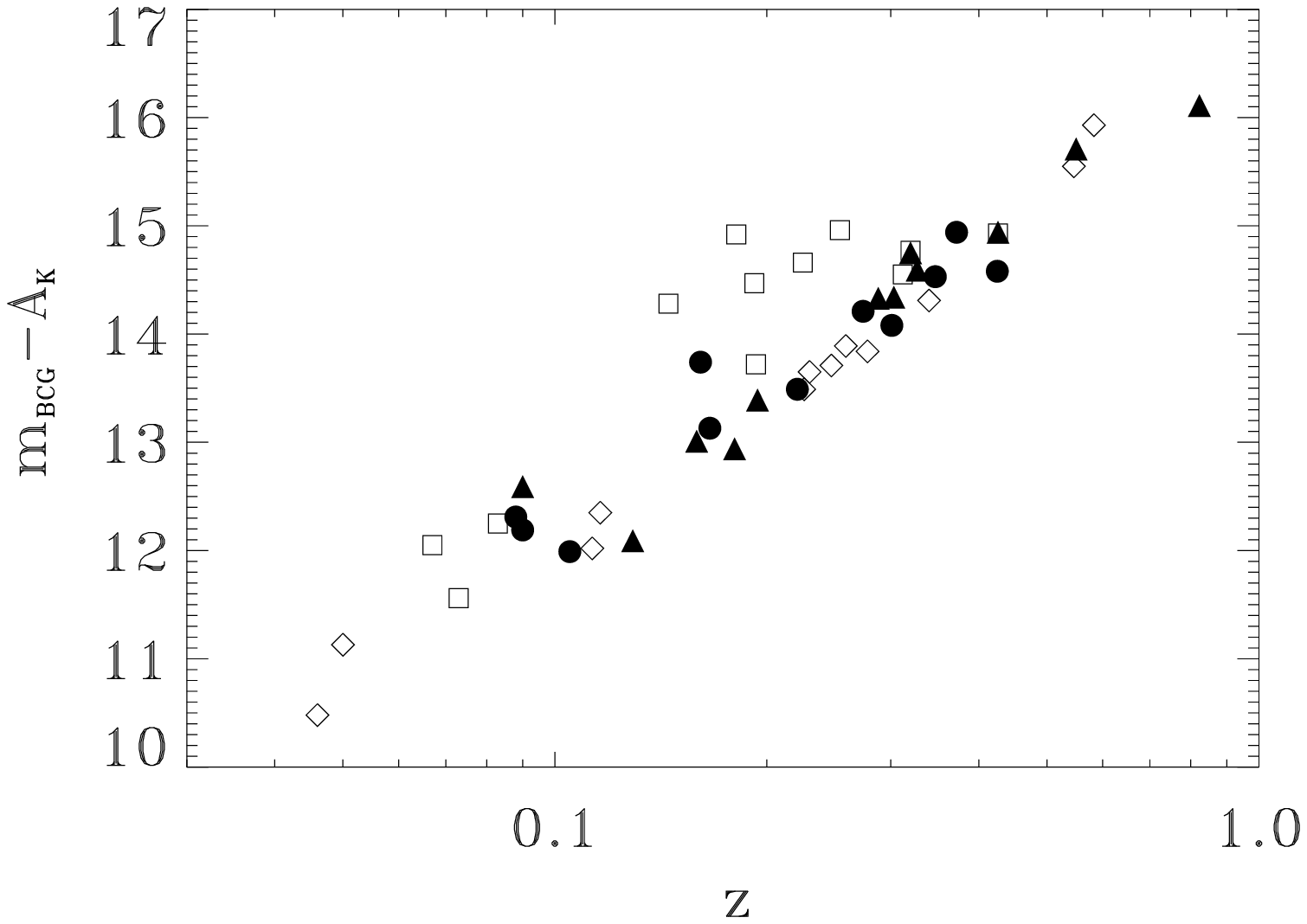,angle=0,width=8.5cm}

	\caption{The K--$z$ diagram from Fig.~\ref{raw_k_z} showing the 
	location of members of quartiles in $\alpha$: the first quartile is
	shown by the squares, the second by the diamonds, the third
	by the triangles and the fourth by the circles.}
 
\label{alphaq}

\end{figure}

\begin{figure}

\epsfig{file=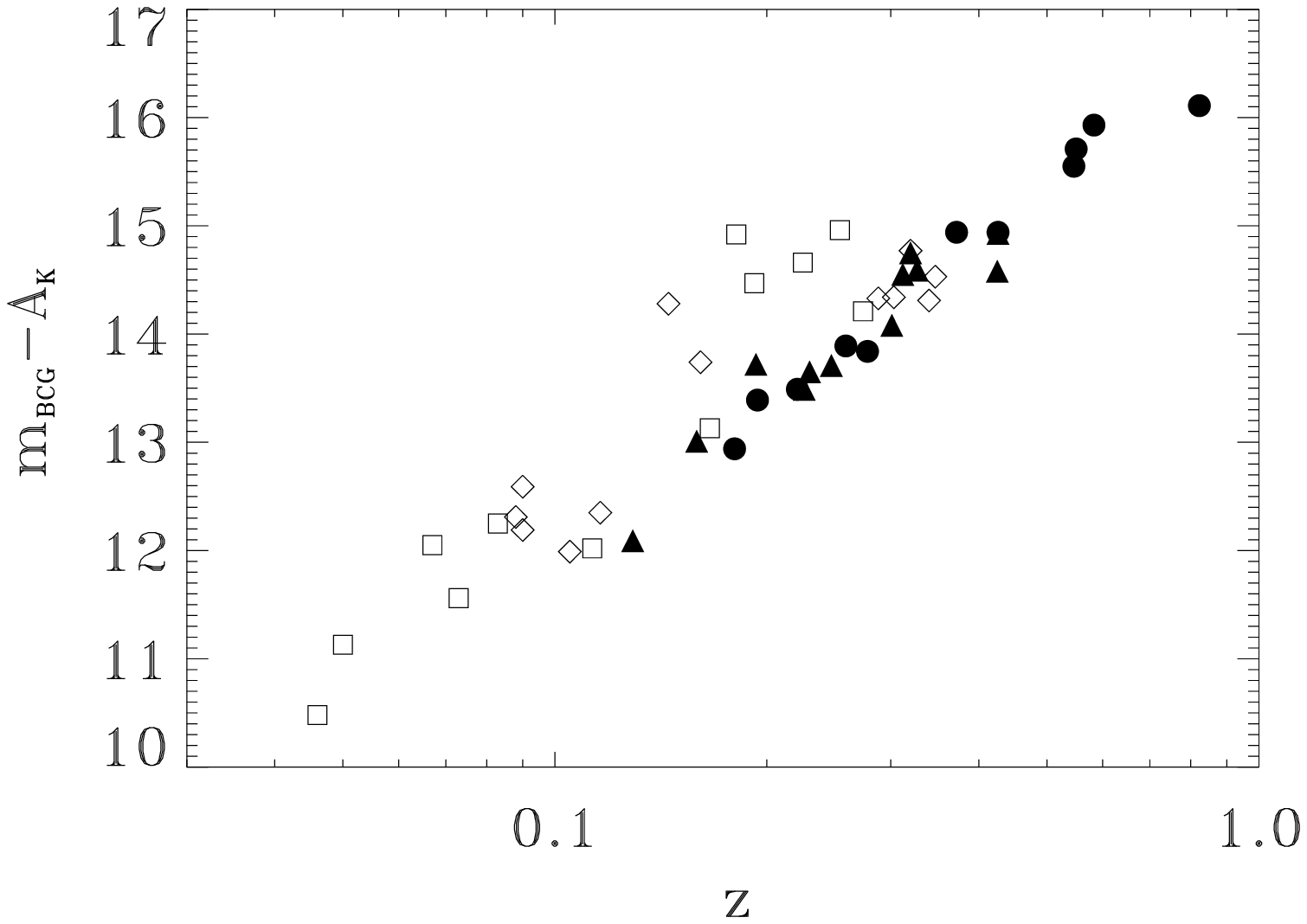,angle=0,width=8.5cm}

	\caption{The K--$z$ diagram from Fig.~\ref{raw_k_z} showing the 
	location of members of quartiles in $L_{\rm X}$: the first quartile is
	shown by the squares, the second by the diamonds, the third
	by the triangles and the fourth by the circles.}
\label{lxq}

\end{figure}

\begin{figure}

\epsfig{file=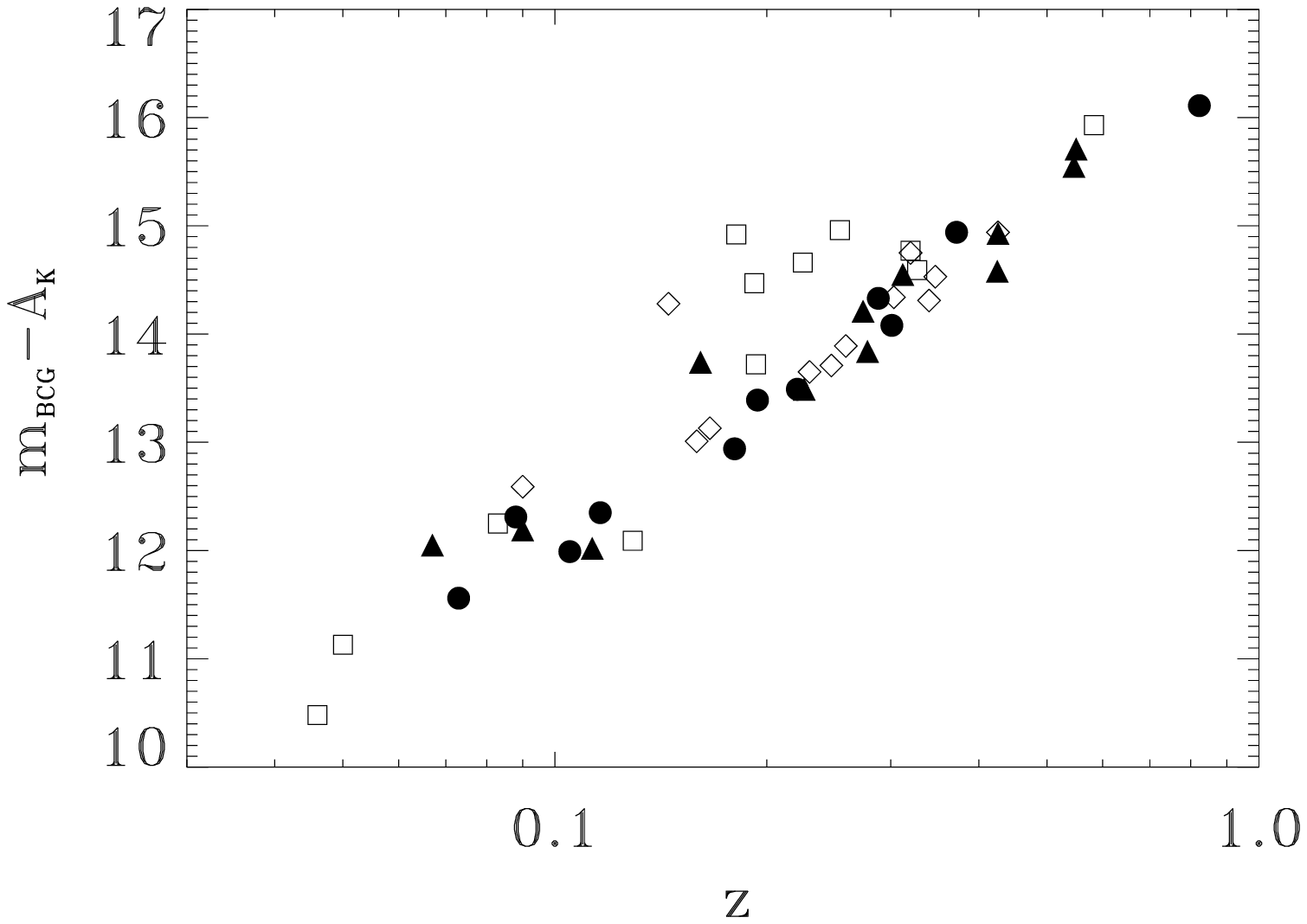,angle=0,width=8.5cm}

	\caption{The K--$z$ diagram from Fig.~\ref{raw_k_z} showing the location
	of members of quartiles in $\hat{N}$: the first quartile is
	shown by the squares, the second by the diamonds, the third
	by the triangles and the fourth by the circles.}

\label{nhatq}

\end{figure}

We then estimated, for each BCG, the number of field galaxies
expected to fall within the 100 kpc counting circle and the 4.5 mag
magnitude strip. This was done using a fit to a compilation of
K band galaxy count data kindly provided by J. Gardner 
(private communication), which comprises data from a number of
authors: Mobasher et al. (1986), Gardner et al. (1993), Glazebrook et
al. (1994), Djorgovski et al. (1995), McLeod et al. (1995),
Metcalfe et al. (1996), Huang et al. (1997) and Gardner et al. (1997).
These data
are shown in Fig.~\ref{kcounts}, together with the following fourth order
polynomial fit to them:
\beq
\log_{10}(\langle N (K)\rangle) \; = \; a_0 + a_1 K +a_2 K^2 + a_3 K^3
+ a_4 K^4,
\label{kcountmodel}
\eeq
over the range $11.25 \la K \la 23.5$,
where $\langle N (K) \rangle$ is the expected galaxy count per
mag per degree$^2$ at magnitude $K$, and the values of the
coefficients in eqn (3) are as follows: 
$a_0=26.509$, $a_1=-8.3590$,
$a_2=0.86876$, $a_3=-3.5824 \times 10^{-2}$ and $a_4=5.2630 \times 10^{-4}$.
This fit has
$\chi^2=128$ for 56 degrees of freedom and this is improved only
marginally by increasing the order of the fit to tenth, say. 
In Table 2 we tabulate, for each BCG, the
raw count $N$ and the corrected count, $\hat{N}$, obtained from it by
subtracting the model background count. In six
cases we find that the estimated contamination exceeds the raw count
$N$: this is not unexpected, given the uncertainty in the background
counts near the limiting magnitudes of the frames (which are,
typically, $20 \la K \la 21$) and our neglect of clustering in the
background population, etc.

\section{CORRELATING BCG AND HOST CLUSTER PROPERTIES}

The correlations between BCG and host cluster properties are graphically
illustrated by Figs.~\ref{alphaq}, \ref{lxq} and \ref{nhatq}, which show, 
on the Hubble diagram of
Fig.~\ref{raw_k_z}, the members of the quartiles of, respectively, $\alpha$,
$L_{\rm X}$ and $\hat{N}$. The similarity between these three
figures shows how all these quantities are broad indicators of 
cluster richness, and, in particular, how the fainter BCGs at $z \la 0.25$ 
tend to be low $\alpha$ galaxies in poor systems, as defined both by the
X-ray luminosity, $L_{\rm X}$, and by the near neighbour number,
$\hat{N}$. This suggests that there would be redundancy in
correcting the $K-z$ relation for all three quantities, so, in this
Section, we perform a non-parametric correlation analysis (using the
Spearman rank correlation coefficient, e.g. Press et al. 1992), to
study the relationship between BCGs and their host clusters and,
hence, to determine the best way to correct the BCG Hubble diagram
for the effects of environment.
 We consider the five properties tabulated in Table 2: (i) redshift, $z$; (ii)
absolute K band magnitude, $M_{\rm K}$, computed using the
K-correction model of equation (1); 
(iii) X-ray luminosity, $L_{\rm X}$; (iv) corrected near neighbour 
number, $\hat{N}$; and (v) BCG structure parameter, $\alpha$.

\begin{table}

\caption{Results of Spearman rank correlation analysis. The upper
half of the table gives the Spearman rank correlation coefficient,
$r_{\rm AB}$, while the lower half gives \mbox{$\hat{P}_{\rm
AB} \equiv-\log_{\rm 10}[P(r_{\rm
AB})]$}, where $P(r_{\rm AB})$ is the probability that the absolute
value of the correlation coefficient would be as large as $|r_{\rm
AB}|$ for a sample of the same size, under the hypothesis that the
quantities $A$ and $B$ are uncorrelated.}

\begin{tabular}{lcccccc}
 & $z$ & $M_{\rm K}$ & $L_{\rm X}$ & $\hat{N}$ & $\alpha$\\ 
$z$    & - & -0.22 & 0.61 & 0.02 & 0.10 \\
 $M_{\rm K}$   & 0.86 & - & -0.49 & -0.39 & -0.40 \\
 $L_{\rm X}$  & 5.36 & 3.29 & - & 0.34 & 0.31 \\
 $\hat{N}$  & 0.04 & 2.15 & 1.74 & - & 0.58 \\
 $\alpha$  & 0.28 & 2.29 & 1.49 & 4.73 & - \\

\end{tabular}

\end{table}  

In Table 3 we present the results of this Spearman rank correlation
analysis: in the upper
half of the table we tabulate the Spearman rank correlation coefficient,
$r_{\rm AB}$, while the lower half gives \mbox{$\hat{P}_{\rm AB}\equiv-\log_{\rm 10}[P(r_{\rm
AB})]$}, where $P(r_{\rm AB})$ is the probability that the absolute
value of the correlation coefficient would be as large as $|r_{\rm
AB}|$ for a sample of the same size, under the hypothesis that the
quantities $A$ and $B$ are uncorrelated. The $P(r_{\rm AB})$ values
are computed under the assumption that the quantity \mbox{$r_{\rm AB} [(N-2)/
(1-r_{\rm AB}^2)]^{1/2}$}, where $N$ is the number of galaxies in
the sample, is distributed like Student's t statistic: from Monte Carlo
simulations we judge this approximation to be very good, typically 
estimating $\hat{P}_{\rm AB}$ to an accuracy of $\sim$0.02.
A very similar set of results to Table 3
is obtained if we use either the constant age K-correction of
Glazebrook et al. (1995)
or the empirical K-correction of Arag\'{o}n--Salamanca et al. (1993), rather 
than the model fit of equation (1).
Note that the use of a rank correlation coefficient underestimates the
true level of correlation between  $M_{\rm K}$ and $\alpha$ to the
extent that this relation is parabolic, as commonly assumed, rather
than monotonic (see Section 7 below): none of the other relations
tabulated in Table 3 are so affected, justifying our use of a rank
correlation analysis to quantify them.

\begin{figure}
	
\epsfig{file=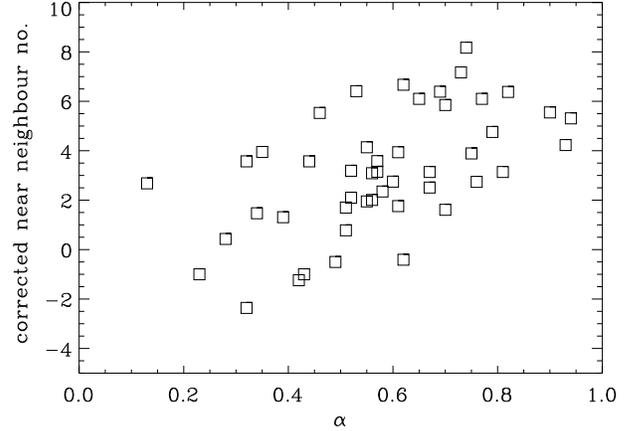,angle=0,width=9cm}

\caption{Correlation of $\hat{N}$ and $\alpha$. This relation is not  
simply the result of contamination within a radius $r_{\rm m}$ of the BCG, 
since these objects were removed before calculating $\alpha$.}
 \label{nhat_alpha}

\end{figure}

A number of these correlations are worthy of comment. Firstly, the
K band absolute magnitudes of our BCGs show no evidence of 
significant evolution with redshift, above the passive evolution 
removed by the K-correction,  but we see from Table 3
that $M_{\rm K}$ is strongly correlated with $L_{\rm X}$, $\hat{N}$
and $\alpha$. The
strong correlation between $z$ and $L_{\rm X}$ simply reflects the 
flux-limited nature of EMSS cluster sample, but note that the strong
correlation between $\hat{N}$ and $\alpha$ (shown in
Fig.~\ref{nhat_alpha}) is {\em not\/} due to near neighbours lying within 
$r_{\rm m}$ of the BCG,
since we explicitly excluded such contamination in calculating
$\alpha$: instead, it illustrates how well $\alpha$ can express the
richness of the BCG's local environment, as argued by Hoessel (1980).

These results could, of course,  be due to correlations with a 
third quantity: for example, we know that $\hat{N}$ and $\alpha$ are
strongly correlated, so the correlation between $\hat{N}$ and $M_{\rm
K}$, say, could just result from that and a correlation between
$M_{\rm K}$ and $\alpha$. To test that we have computed,  for our
five physical properties, partial
Spearman correlation coefficients (e.g. Yates et al. 1986) of the form
$r_{\rm AB,C}$, where
\beq
r_{\rm AB,C} \; = \; \frac {r_{\rm AB} -  r_{\rm AC} r_{\rm BC}}
{[(1-r^2_{\rm AC})(1-r^2_{\rm BC})]^{1/2}},
\eeq
where $r_{\rm AB}$, etc, are the Spearman rank correlation
coefficients, as before. The significance of the partial rank
correlation coefficients may be calculated using a similar
approximation as above -- that \mbox{$r_{\rm AB,C} [(N-3)/
(1-r_{\rm AB,C}^2)]^{1/2}$} is distributed as Student's t statistic --
to compute the probability, $P(r_{\rm AB,C})$, of obtaining 
a partial rank correlation coefficient with absolute value as large
as $|r_{\rm AB,C}|$ under the null hypothesis that the correlation
between $A$ and $B$ results solely from correlations between $A$ and
$C$ and between $B$ and $C$: Monte Carlo simulations show that this
approximation works as well for  $P(r_{\rm AB,C})$ as for  $P(r_{\rm AB})$.

We relegate the full set of partial rank correlation results to an Appendix,
and note only a few important points here. 
The most striking result is that once the dependence on X-ray
luminosity has been removed, $\alpha$ becomes independent of $z$ to
a high degree: $P(r_{\rm z\alpha,L_{\rm X}})=0.60$.
We find that the
$\hat{N}-\alpha$ relation is not due to separate correlations between
those two variables and a third: the lowest $P(r_{\rm AB,C})$ for
$A=\hat{N}$ and $B=\alpha$ arises when $C=M_{\rm K}$, but, even then,
there is a probability of less than $4\times10^{-4}$ that the
$\hat{N}-\alpha$ correlation is entirely due to the $M_{\rm K}-\hat{N}$ and
$M_{\rm K}-\alpha$ correlations. This leads to high
values of $P(r_{\rm AB,C})$ when $A=\hat{N}$ and $C=\alpha$, or vice
versa, since these variables are almost interchangable. There is some
reduction in the significance of the $M_{\rm K}-L_{\rm X}$ correlation
when it is evaluated at constant $\alpha$, but it remains quite
strong, consistent with the suggestion by Hudson \& Ebeling (1997)
that correction for  $L_{\rm X}$ can remove some of the residual 
environmental dependence of BCG magnitudes left after the 
$L_{\rm m}-\alpha$ correction has been made.
Finally, it appears that none of the significant correlations
between pairs of properties arise from separate evolutionary trends of
the properties with redshift. 
To summarise, it is clear that $M_{\rm K}$ is strongly correlated with
both $\alpha$ and $L_{\rm X}$, and that no further correction for $\hat{N}$
need be made after the $\alpha$-correction, due to the strength of the
correlation between $\hat{N}$ and $\alpha$. 

\section{THE CORRECTED K-$\bmath{\lowercase{z}}$ RELATION}

\begin{figure}

\epsfig{file=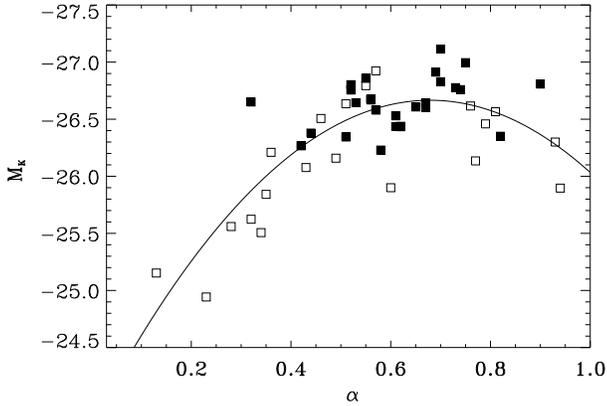,angle=0,width=8.5cm}

\caption{The $L_{\rm m}-\alpha$ relation for our data set. The
	filled and empty squares denote the BCGs in clusters with
	X-ray luminosities greater than and less than $2.3 \times 
	10^{44}$ erg s$^{-1}$ respectively. The line shows the
	best quadratic fit to the data.}
 \label{mk_alpha}

\end{figure}

The relationship between absolute magnitude and $\alpha$ for our 
BCG sample is shown in Fig.~\ref{mk_alpha}. The relationship diverges from a
purely linear slope at large values of $\alpha$ and is in this respect
qualitatively similar to those of  Hoessel (1980), who used a sample of 108
Abell clusters, and Postman \& Lauer (1995), who determined $\alpha$ for a
sample of 119 BCGs at $z\leq0.05$ in the R band. In a similar way to
Postman \& Lauer (1995), we fit  the correlation between
structure parameter and K band luminosity with a parabola, which gives a
least-squares best fit of
\beq
M_{\rm K}=-23.84 - (8.34\pm1.14) \alpha + (6.14\pm1.00) \alpha^2,
\label{mk_alpha_fit}
\eeq
which is also shown in Fig.~\ref{mk_alpha}. The vertical scatter about
this fit is 
$\sigma=0.29$ mag, and the gradient in the linear
part  of the relation is significantly  steeper for our sample than
for the R band sample of Postman \& Lauer (1995). 
It is clear that a parabolic fit is a reasonable description of the
data over the range $0.1\leq\alpha\leq1.0$, and Fig.~\ref{mk_alpha}
provides the first real evidence that the $L_{\rm m}-\alpha$ relation
at least flattens at high $\alpha$, rather than
continuing linearly with increased scatter, as appears the case for
the Postman \& Lauer (1995) data. 

\begin{figure}

\epsfig{file=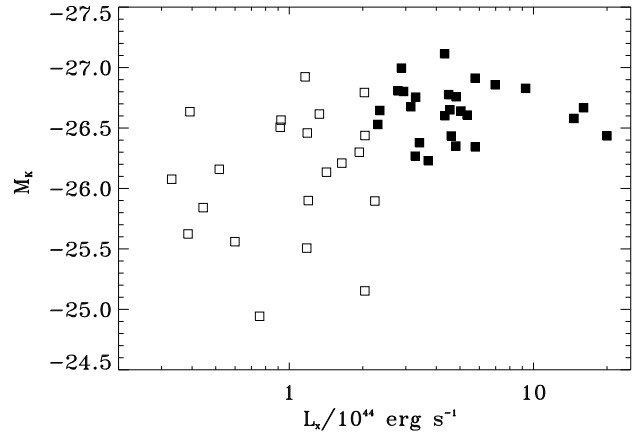,angle=0,width=9cm}

	\caption{The relationship between X-ray luminosity and
	raw BCG absolute magnitude. The empty squares denote the
	BCGs in clusters with $L_{\rm X} < 2.3 \times 10^{44}$
	erg s$^{-1}$, while the filled squares are the BCGs
	in clusters above that threshold. Note the much wider
	spread of absolute magnitudes seen for the BCGs in the
	less X-ray-luminous clusters.}
 \label{mk_lx}

\end{figure}

\begin{figure}

\epsfig{file=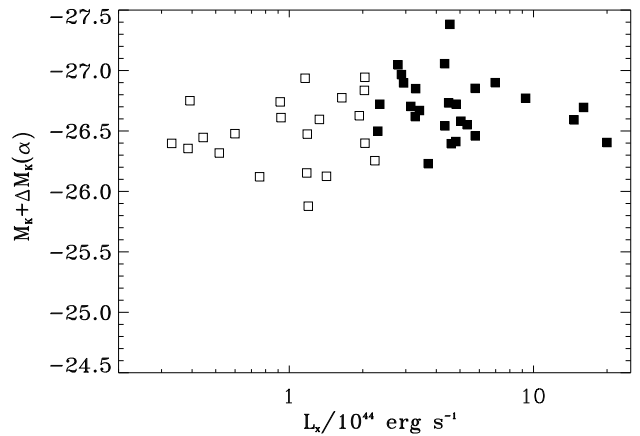,angle=0,width=9cm}

	\caption{The relationship between X-ray luminosity and
	$\alpha$-corrected BCG absolute magnitude. The empty squares 
	denote the
	BCGs in clusters with $L_{\rm X} < 2.3 \times 10^{44}$
	erg s$^{-1}$, while the filled squares are the BCGs
	in clusters above that threshold. Comparison of this
	plot and Fig.~\ref{mk_lx} shows how the $\alpha$ correction
	removes most of the scatter in the absolute magnitudes
	of the low-$L_{\rm X}$ BCGs, while leaving the 
	high-$L_{\rm X}$ ones virtually unchanged. }

 \label{mkalphacor_lx}

\end{figure}

The filled squares in Fig.~\ref{mk_alpha} are the BCGs in high-$L_{\rm X}$
clusters. Note that these are all to be found near the peak of
the parabola, indicating how high-$L_{\rm X}$ selection picks
out BCGs with a narrow range of absolute magnitudes, and 
suggesting that these magnitudes will be largely unaffected by 
the correcting for the $M_{\rm K}-\alpha$ relation of equation 
(5). This is graphically demonstrated by Figs.~\ref{mk_lx} and 
\ref{mkalphacor_lx}, which show,
respectively, the relationship between $L_{\rm X}$ and the raw
and $\alpha$-corrected $M_{\rm K}$ values. 
Below host cluster X-ray luminosities of $L_{\rm X}\simeq2.3 \times
10^{44}$ erg s$^{-1}$, there is a wide dispersion in raw BCG
absolute magnitudes, which is largely, but not completely, removed
by the $M_{\rm K}-\alpha$ correction, while the high-$L_{\rm X}$
absolute magnitudes are barely changed. In what follows we use this
threshold of $L_{\rm X}=2.3 \times 10^{44}$ erg s$^{-1}$ to mark the
break between a homogeneous population of BCGs in high-$L_{\rm X}$
and a low-$L_{\rm X}$ population which displays a wider scatter in its
properties. Intriguingly, the $L_{\rm X}$ threshold
we empirically found on the basis of BCG properties is almost identical 
to that Annis (1997) has recently argued marks the dividing point in the
X-ray properties of EMSS clusters: above $2 \times
10^{44}$ erg s$^{-1}$ the X-ray properties are very homogeneous, while
below that, Annis detects two populations of clusters, with quite
differing properties, one of which he argues contains clusters where the X-ray
emission is solely from the intracluster medium (the only sort of
cluster found above  $2 \times 10^{44}$ erg s$^{-1}$) and the
other where a significant contribution comes from a cooling flow. We note 
that the lack of strong correlation between $M_{\rm K}$ and $\alpha$ for 
high $L_{\rm X}$ clusters found here, is in contrast to the results 
found by Hudson \& Ebeling (1997), who argue that the optical 
$L_{\rm m}-\alpha$ 
relation actually steepens for high-$L_{\rm X}$ Abell clusters. However, 
their local Abell sample does not probe  luminosities as high as the EMSS 
clusters -- only 2 clusters in their sample have $L_{\rm X}\geq2.3 
\times 10^{44}$ erg s$^{-1}$ in the EMSS passband.

After correction for the $M_{\rm K}-\alpha$ relation of equation (5)
we find that the rms dispersion in absolute magnitudes falls from
$\sigma=0.47$ mag to $\sigma=0.29$ magnitudes, while the dispersion
for the high-$L_{\rm X}$ subsample (the 26 galaxies in clusters with
$L_{\rm X} > 2.3 \times 10^{44}$ erg s$^{-1}$) about its own
best-fit $M_{\rm K}-\alpha$ relation is 0.22 mag, identical to the rms
dispersion in their raw magnitudes: the best-fit relation
for the high-$L_{\rm X}$ sample is $M_{\rm K}=-26.18-(0.87\pm2.34)\alpha
-(0.18\pm1.89) \alpha^2$, indicating the weakness of the dependence of
$M_{\rm K}$ on $\alpha$ shown in Fig.~\ref{mk_alpha} for these BCGs.

The residuals about the best $M_{\rm K}-\alpha$ fit for the 
high-$L_{\rm X}$ sample do not correlate with $z$, $L_{\rm X}$ or
$\hat{N}$, while, for the full sample, the only significant correlation 
is with $L_{\rm X}$, similar to that reported by Hudson \& Ebeling
(1997). We also tried correcting our raw absolute magnitudes
using the same form of $L_{\rm m}-\alpha-L_{\rm X}$ relation recently
advocated by Hudson \& Ebeling (1997) which, translated to the K band,
reads
\begin{eqnarray}
M_{\rm K} &  = & c_0 + c_1 \alpha + c_2 \alpha^2 + c_3
\log_{10}(\hat{L}_{\rm X}) + c_4 \alpha \log_{10}(\hat{L}_{\rm X})
\nonumber \\
& + & c_5 \alpha^2 \log_{10}(\hat{L}_{\rm X}),
\end{eqnarray}
where $\hat{L}_{\rm X}$ is the cluster X-ray luminosity in units of
$10^{44}$ erg s$^{-1}$. For the full sample, the best fit values of
these parameters are $c_0=-23.26$, $c_1=-10.66\pm1.58$,
$c_2=8.45\pm1.49$, $c_3=-3.63\pm1.13$, $c_4=12.23\pm4.30$ and 
$c_5=-10.37\pm3.97$, which means that the terms are in similar
proportion to those found by Hudson \& Ebeling (1997) for the
Lauer \& Postman (1994) optical BCG sample. With this model, the
rms dispersion of the full sample is decreased slightly to 0.26 mag,
while that for the high-$L_{\rm X}$ sample actually increases slightly
to 0.23 mag, as the loss of degrees of freedom arising from the use 
of three more parameters in the fit outweighs the marginal decrease
in the scatter in absolute magnitudes brought about by correction for
the residual dependence on $L_{\rm X}$ left after correction for $\alpha$.

In principle, galaxy merging in the centre of clusters should increase the 
stellar mass of the BCGs with time. Hausman \& Ostriker (1978) show that
the $\alpha$ parameter is a sensitive measure of the merging history of
BCGs: therefore, we can determine the effect
that the merging process has had on our sample by using the $\alpha$ parameter 
as a tracer of the growth of the BCGs through merging. As the Spearman results 
of Section 6 for the full sample show, once the dependences of $\alpha$ on 
$L_{\rm X}$ and of $L_{\rm X}$ on $z$ are removed, $\alpha$ is 
independent of $z$: there is a 60 per cent probability that the correlation
between $\alpha$ and $z$ arises from the combination of their individual 
correlations with $L_{\rm X}$. This lack of a correlation between $\alpha$
and $z$ is slightly more strongly confirmed for the high-$L_{\rm X}$ 
sample of 26 BCGs, as the Spearman test for a correlation 
between $\alpha$ and $z$ for these galaxies gives a
correlation coefficient with probability of 0.63 of arising
from uncorrelated variables. We conclude, therefore, that the
BCGs in the high-$L_{\rm X}$ clusters sample form  a very homogeneous
population, with an intrinsic dispersion in absolute magnitude of
no more than 0.22 mag, and that there has been no significant 
evolution due to galaxy merging since a redshift of $z\sim0.8$. It follows 
that our high-$L_{\rm X}$ BCG sample constitute a set of standard candles
ideal for use in classical cosmological tests, such as the
estimation of  $\Omega_{\rm M}$ and $\Omega_\Lambda$.

\section{ESTIMATING $\Omega_{\rm M}$ AND $\Omega_\Lambda$ FROM THE BCG
HUBBLE DIAGRAM}

The Hubble diagram for a set of standard candles determines the relationship
between luminosity distance, $d_{\rm L}$, and redshift, which may be
written (Carroll, Press \& Turner 1992; Perlmutter et al. 1997) in the
general form 
\begin{eqnarray}
&d_{\rm L} (z,\Omega_{\rm M},\Omega_\Lambda,H_0)  =  \frac{c(1+z)}
{H_0\sqrt{|\kappa}|} \times \nonumber \\
 \!\!{\cal I}& \!\! \!\! \!\!\left\{ \!\!\sqrt{|\kappa|} \int_0^z \!\!dx
\left[(1+x)^2(1+\Omega_{\rm M}x)-x(2+x)\Omega_\Lambda\right]^{-1/2} 
\right\}
\label{dl}
\end{eqnarray}
where: (i) if $\Omega_{\rm M}+\Omega_\Lambda > 1$, then
${\cal I}(x)=\sin(x)$ and $\kappa=1-\Omega_{\rm M}-\Omega_\Lambda$;
(ii) if  $\Omega_{\rm M}+\Omega_\Lambda < 1$, then
${\cal I}(x)=\sinh(x)$ and $\kappa=1-\Omega_{\rm M}-\Omega_\Lambda$;
and (iii) if  $\Omega_{\rm M}+\Omega_\Lambda=1$, then ${\cal I}(x)=x$
and $\kappa=1$. As emphasised by Perlmutter et al. (1997), it is
important to note that the form of equation (7) means that
$d_{\rm L}$ depends on  $\Omega_{\rm M}$ and $\Omega_\Lambda$
independently, and not solely through the deceleration parameter,
\mbox{$q_0 \equiv \Omega_{\rm M}/2 - \Omega_\Lambda$} -- i.e. the confidence
region in the  $(\Omega_{\rm M},\Omega_\Lambda)$ plane determined
from the Hubble diagram of a set of standard candles is not parallel
to contours of constant $q_0$, except at $z \ll 1$. We prefer,
therefore, to solve for  $\Omega_{\rm M}$ and $\Omega_\Lambda$ using
our Hubble diagram for BCGs in high-$L_{\rm X}$ clusters, rather than 
for $q_0$. 

We do that, using the following procedure for each candidate
cosmology:

\begin{enumerate}

\item Select values for $\Omega_{\rm M}$ and $\Omega_\Lambda$: $h=0.5$
is assumed throughout.

\item Using equation (7), compute the angle, $\theta_{\rm
m}$, corresponding to a physical distance  $r_{\rm m}=25$ kpc at
the redshift of each of the 47 BCGs:
$\theta_{\rm m}=r_{\rm m}/d_{\rm A}$, where $d_{\rm A}=d_{\rm
L}/(1+z)^2$ is the angular diameter distance.

\item Calculate the change in the apparent magnitudes of the BCGs,
as \mbox{$\Delta m_{\rm BCG}=\alpha(d_{\rm A}-d_{\rm A,ref})/d_{\rm A,ref}$},
where the $\alpha$ values are those in Table 2, which assume 
$\Omega_{\rm M}=1$ and $\Omega_\Lambda=0$, and $d_{\rm A,ref}$
is the angular diameter distance in that reference model. This first
order correction to the apparent magnitude should be accurate so long
as $r_{\rm m}$ does not vary much from the reference aperture with
$\Omega_{\rm M}=1$ and $\Omega_\Lambda=0$, and that condition holds
true for all physically reasonable cosmologies.

\item Compute the ages of the galaxies in the cosmology, assuming their
formation at a redshift $z_{\rm f}=3$, and compute their
K--corrections using the Bruzual \& Charlot models, as described in
Section 4. Use the K-corrections for, and luminosity distances of, the
galaxies in the cosmology to derive the BCG absolute magnitudes.

\item Perform a quadratic fit to the new $M_{\rm K}-\alpha$ relation,
and use it to correct the BCG magnitudes to a fiducial $\alpha$ value,
which is taken to be the median $\alpha$.

\item Take  the mean of the $\alpha$-corrected $M_{\rm K}$ as the
estimate for $M_0$, the BCG K band absolute magnitude, and deduce
its standard error, from the scatter of the $M_{\rm K}$ values
about $M_0$.

\item For the redshifts of the 26 BCGs in high-$L_{\rm X}$ clusters,
calculate the predicted apparent magnitude, given $M_0$ and the
$K(z)$ relation in the cosmology. Using the errors on the measured
magnitudes of the  26 BCGs in high-$L_{\rm X}$ clusters, and the
standard error on  $M_0$, compute $\chi^2$ for the comparison of the
26 BCG apparent magnitudes and the apparent magnitudes predicted at
their redshifts in the cosmology.

\end{enumerate}

We report these results for two classes of cosmology, with and without
a cosmological constant.

\subsection{Friedmann models: $\Omega_\Lambda=0$}

\begin{figure}

\epsfig{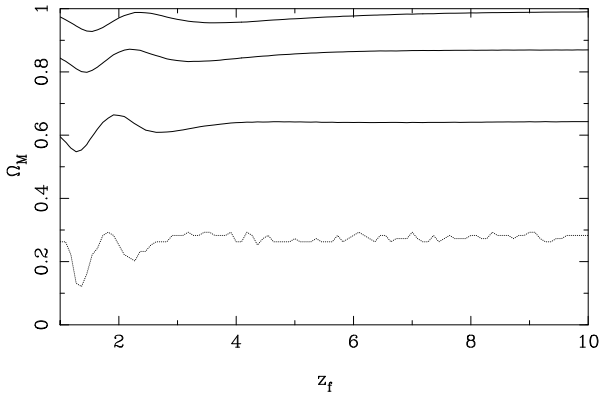}

\caption{The joint constraints on the value of $\Omega_{\rm M}$ 
and the BCG formation redshift $z_{\rm f}$ deduced from the Hubble diagram
for the BCGs in our high--$L_{\rm X}$ clusters. The dotted
line shows the locus of the best-fit $\Omega_{\rm M}$ value for one hundred
values of $z_{\rm f}$ in the range \mbox{$1 \leq z_{\rm f} \leq 10$}, while the
solid contours mark the 68, 95 and 99 per cent confidence contours for
$\Omega_{\rm M}$ and $z_{\rm f}$: the small amplitude random fluctuations in
the dotted line are numerical artefacts introduced by our method of 
estimating $\Omega_{\rm M}$,
while the behaviour of all the curves in the range 
$1.25 \leq z \leq 2.25$ is caused by the transient presence of
light from AGB stars in the stellar synthesis model used to compute
the BCG K-corrections, as discussed in Section 4.}

 \label{omega_zf}

\end{figure}

Much of the previous work trying to estimate cosmological parameters 
from  the Hubble diagrams of distant
galaxies has assumed $\Omega_\Lambda=0$, in which case the
deceleration parameter, $q_0$, is given by $q_0=\Omega_{\rm M}/2$.  
Table 4 summarises the $q_0$ values obtained previously, 
from  a variety of galaxy samples, using
BCGs and radio galaxies, in both the optical and the near-infrared.
Several points are worthy of note, motivating the use of K band
photometry of BCGs in X-ray clusters to estimate $q_0$. Firstly,
as discussed above, the use of K band photometry is to be preferred
to the use of optical photometry, due to the insensitivity of K band
light to star formation history of the galaxy. This means that our
estimated $q_0$ value is insensitive to uncertainties in the age of
the stellar population of the BCGs and in the stellar IMF: optical
photometry would produce a greater uncertainty in $q_0$ on both those
scores. Secondly, X-ray cluster selection is to be preferred over
optical cluster selection, since, as we have shown, it is very
effective at selecting galaxies with a very narrow intrinsic
dispersion in absolute magnitude. Thirdly, BCGs are to be preferred
over radio galaxies, both because the intrinsic dispersion in radio
galaxy absolute magnitudes appears larger ($\sim 0.5$ mag) and 
because bias may be introduced by the significant non-stellar
component to the K band light (Dunlop \& Peacock 1993; Eales et al.
1997).

If we assume $\Omega_\Lambda=0$, then we obtain 
\mbox{$\Omega_{\rm M}=0.28\pm0.24$} if $z_{\rm
f}=3$. Consistent values are obtained if $z_{\rm f}$ is varied between
$z_{\rm f}=1$ (i.e. slightly higher than the redshift of our most 
distant BCG) to $z_{\rm f}=10$, as shown in Fig.~\ref{omega_zf}.
These results are consistent 
with the results of Perlmutter et al. (1997), who find \mbox{$\Omega_{\rm M}
=0.88^{+0.69}_{-0.60}$} for a  $\Omega_\Lambda=0$ cosmology, from
the Hubble diagram of the first seven high-$z$ Type Ia supernovae
discovered by  The Supernova Cosmology Project (although our best-fit
$\Omega_{\rm M}$ value is clearly lower than theirs, and our
confidence interval is much narrower) and with those of Carlberg et
al. (1996), who find
$\Omega_{\rm M}\simeq0.24\pm0.09$ from CNOC cluster mass profiles.

\begin{table*}

\caption{Previous estimates of the deceleration parameter $q_0$ from the
Hubble diagram of BCGs and radio galaxies. The symbols in the waveband 
column denote whether the measurements were carried out in the optical (O)
or near-infrared (IR). Wherever possible the $q_0$ values listed are those 
quoted by the authors after correcting for evolution, otherwise the raw 
values are given.}

\begin{tabular}{|c|c|c|c|c|l|}

Sample & $z$ range & Waveband & Number & $q_0$ & Reference \\ \hline
BCG & $\leq0.2$ & O & 17 & $2.5\pm1.0$ & Humason, Mayall \& Sandage (1956)\\
BCG & $\leq0.46$ & O & 8 & $1.0\pm0.5$ & Baum (1957, 1961a,b) \\
BCG & $\leq0.46$ & O & 38 & $1.5\pm0.9$ & Peach (1970) \\
BCG & $0.003-0.46$ & O & 84 & $0.96\pm 0.4$ & Sandage (1972b) \\
BCG & $0.005-0.46$ & O & 98 & $1 \pm1$ & Sandage \& Hardy (1973) \\
BCG & $0.01-0.47$ & O & 68 & $0.33$ or $-1.27$ & Gunn \& Oke (1975) \\
BCG & $0.04-0.46$ & O & 67 & $1.0\pm0.3$ & Sandage, Kristian \& Westphal (1976) \\
BCG & $0.04-0.75$ & O & 50 & $1.6\pm0.4$ & Kristian, Sandage \& Westphal (1978)\\
radio & $0.14-0.95$ & IR & 6 & $\simeq 0$ & Lebofsky (1980) \\
radio & $0.03-1.6$ & IR & 83 & $0-0.6$ & Lilly \& Longair (1984) \\
radio & $0.016-0.9$ & IR & 61 & $\simeq 0.5$ & Lebofsky \& Eisenhardt (1986)\\
BCG & $0.017-0.147$ & O & 116 & $-0.55\pm0.45$ & Hoessel, Gunn \& Thuan (1980) \\
BCG & $0.15-0.83$ & O & 41 & $0.3\pm(0.21-0.39)$ & Clowe, Luppino \& Gioia (1995) \\
X-ray BCG & $0.11-0.83$ & IR & 26 & $0.14\pm0.12$ & this work\\
\end{tabular}

\end{table*}

\subsection{Friedmann-Lema\^{\i}tre models:  $\Omega_\Lambda \neq 0$}

\begin{figure}

\epsfig{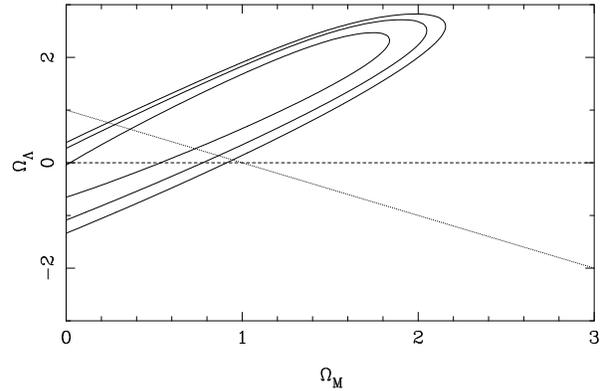}

\caption{The joint constraints on $\Omega_{\rm M}$ and
$\Omega_\Lambda$ obtainable from the Hubble diagram of the 26 BCGs
in high-$L_{\rm X}$ clusters. The contours mark the 68, 90 and 95
per cent confidence levels for  $\Omega_{\rm M}$ and $\Omega_\Lambda$
jointly, while the the dashed and dotted lines mark the models with
$\Omega_\Lambda=0$ and $\Omega_{\rm M}+\Omega_\Lambda=1$ respectively.}

 \label{joint}

\end{figure}

Fig.~\ref{joint} shows the joint constraints on $\Omega_{\rm M}$
and $\Omega_{\Lambda}$ we obtain if we allow a non-zero cosmological
constant: this is very similar to the corresponding plot presented
by Perlmutter et al. (1997).
If we consider only spatially flat models, where 
\mbox{$\Omega_{\rm M}+\Omega_\Lambda=1$}, we find
\mbox{$\Omega_{\rm
M}=0.55^{+0.14}_{-0.15}$}, with a corresponding 95 per cent upper
limit to $\Omega_\Lambda$ of 0.73, for $z_{\rm f}=3$: as above, these 
figures are very stable, varying marginally over the full range
\mbox{$z_{\rm f} \leq 10$}.
Perlmutter et al. (1997) find \mbox{$\Omega_{\rm
M}=0.94^{+0.34}_{-0.28}$} if they assume a spatially flat Universe,
which is again consistent with our results:
their results translate into a 95 per cent confidence level of
$\Omega_\Lambda < 0.51$, slightly tighter than ours. 
From the incidence of gravitational lensing of quasars, 
Kochanek (1996) derived a 95 per cent limit of $\Omega_\Lambda <
0.66$, while two recent studies have reported significant detections
of non-zero $\Omega_\Lambda$ values: 
Henry (1997) finds $\Omega_{\rm M}=0.55\pm0.17$ from the
evolution of the cluster X-ray temperature function, and Bender
et al. (1997) deduce  $\Omega_{\rm M}\sim0.5$, $\Omega_\Lambda\sim0.5$
from cluster elliptical Fundamental Plane relations.

\section{DISCUSSION AND CONCLUSIONS}

This paper has presented the K band Hubble diagram for a sample of
BCGs in X-ray selected clusters. Our principal results are that the 
brightest X-ray clusters contain BCGs with a very
narrow spread of absolute magnitudes, and that this, coupled with
the insensitivity of K band light to the details of the star formation
processes in, and histories of, galaxies, makes samples such as ours
ideal for performing classical cosmological tests. We find that the BCGs in 
the clusters with high X-ray luminosities ($L_{\rm X} > 2.3 
\times 10^{44}$ erg
s$^{-1}$ in the 0.3  - 3.5 keV passband) have an rms dispersion in
absolute K band magnitude of no more than 0.22 mag, and that leads
us to an estimate of \mbox{$\Omega_{\rm M}=0.28\pm0.24$} 
if $\Lambda=0$, while, for a spatially flat Universe, we find
$\Omega_{\rm M}=0.55\pm0.15$ and a 95 per cent limit of 
\mbox{$\Omega_\Lambda < 0.73$}, 
with  only a very weak sensitivity to the properties of the 
stellar population of the BCGs over a wide range of redshifts.
The stability of these results demonstrates that BCGs are excellent
standard candles, competitive with Type Ia supernovae (which also
have an intrinsic magnitude dispersion of $\sim0.2$ mag: Perlmutter
et al. 1997) in the
estimation of cosmological parameters $\Omega_{\rm M}$ and 
$\Omega_\Lambda$: the further refinement of these
two techniques over the next few years holds out hope for the 
determination of these fundamental cosmological parameters with
high precision.

Finally, the lack of evolution we see in the K band, shown by  the
absence of a correlation between the BCG structure parameter,
$\alpha$, and redshift, adds to the growing body of evidence that
cluster ellipticals are old systems, possibly undergoing only
quiescent evolution of their stellar populations (e.g. Dunlop et al.
1996; Bender et al. 1996; Ellis et al. 1997). It is intriguing to note
that a qualitatively similar lack
of evolution is seen in the X-ray luminosity (Collins et al. 1997,
Burke et al. 1997) and temperature (Mushotsky \& Scharf 1997, Henry
1997) functions. We shall return to this question of the relationship
between the formation and evolution of BCGs and their host clusters in
a forthcoming paper, incorporating recently acquired K band imaging
data for BCGs in clusters from the deep Southern SHARC {\em ROSAT\/} cluster
survey (Collins et al. 1997): these clusters were selected above a
flux limit an order of magnitude fainter than that of the EMSS,
enabling us to disentangle the effects of redshift and X-ray
luminosity, in a manner impossible with the current data set.

\section*{ACKNOWLEDGMENTS} 

CAC acknowledges support from a  PPARC Advanced Fellowship and RGM 
that from PPARC
rolling grants at QMW and Imperial College. We thank Colin Aspin,
John Davies, Dolores Walther and Thor Wold for their support during the
two UKIRT runs, Jon Gardner for making available his compilation of
K-band galaxy count data, Lance Miller, Andy Fabian and Doug Burke for useful
discussions,  and  are grateful to the referee, Mike Hudson,
for suggesting changes which improved the paper. The United Kingdom
Infrared Telescope
 is operated by the Joint Astronomy Centre on behalf of the
U.K.  Particle Physics and Astronomy Research Council.

\section*{REFERENCES}

\myref{Abell G.O., 1958, ApJS, 3, 211}
\myref{Allen S.W., 1995, MNRAS, 276, 947}
\myref{Andreon S., Garilli B., Maccagni D., Vettolani G., 1992, AA,
	266, 127}
\myref{Annis J., 1997, preprint}
\myref{Arag\'{o}n-Salamanca A., Ellis R., Couch W.J., Carter D., 1993,
	MNRAS, 262, 794}
\myref{Baum W.A., 1957, AJ, 62, 6}
\myref{Baum W.A., 1961a, Observatory, 81, 114}
\myref{Baum W.A., 1961b, in McVittie G.C., ed, Proc. IAU
	Symp. 15. MacMillan, New York, p. 390}
\myref{Bautz L.P., Morgan W.W., 1970, ApJ, 162, L149}
\myref{Bender R., Saglia, R.P., Ziegler, B., Belloni, P., Greggio, L.,
	Hopp, U., 1997, ApJ, in press}
\myref{Bender R., Ziegler B., Bruzual G., 1996, ApJ, 463, L51}
\myref{Bershady M.A., Hereld M., Kron R.G., Koo D.C., Munn J.A., Majewski
S.R., 1994, AJ, 108, 870}
\myref{Bershady M.A., 1995, AJ, 109, 87}
\myref{Bruzual A.G., Charlot S., 1993, ApJ, 405, 538}
\myref{Burke D.J., Collins, C.A., Sharples, R.M., Romer, A.K., Holden,
	B.P., Nichol R.C., 1997, ApJ, 488, L83}
\myref{Carlberg R.G., et al., 1996, ApJ, 462, 32}
\myref{Carroll S.M., Press, W.H., Turner, E.L., 1992, ARAA, 30, 499}
\myref{Casali M.M., Hawarden T.G., 1992, JCMT-UKIRT Newsletter, no. 3, 33}
\myref{Charlot S., Worthey G., Bressan A., 1996, ApJ, 457, 625}
\myref{Clowe D., Luppino G., Gioia I.M, 1995, in Trimble V., Reisenegger
	A., ed, Clusters, Lensing, and the Future of the Universe, Vol. 88}
\myref{Collins C.A., Burke D.J., Romer A.K., Sharples R.M., Nichol R.C.,
	1997, ApJ, 479, L117}
\myref{Djorgovski S.G., Soifer B.T., Pahre M.A., Larkin J.E., Smith
	J.D., Neugebauer G., Smail I., Matthews K., Hogg D.W., 
	Blandford R.D., Cohen J., Harrison W., Nelson J., 1995, ApJ,
	438, L13}
\myref{Draper P.W., Eaton N., 1996, {\sc pisa}, Starlink User Note
	109, {\tt http://star-www.rl.ac.uk/star/docs/sun109.htx 
	/sun109.html}}
\myref{Dunlop J., Peacock J. A., 1993, MNRAS, 263, 936}
\myref{Dunlop J., Peacock J., Spinrad H., Dey A., Jimenez R., Stern D.,
	Windhorst, R., 1996, Nature, 381, 581}  
\myref{Eales S., et al., 1997, MNRAS, in press} 
\myref{Edge A.C., 1991, MNRAS, 250, 191}
\myref{Ellis R.S. Smail I., Dressler A., Couch W.J., Oemler A.,
Butcher H., Sharples R.M., 1997, ApJ, 483, 582} 
\myref{Gardner J.P., Cowie L.L., Wainscoat R.J., 1993, ApJ, 415, L9}
\myref{Gardner J.P., Sharples R.M., Carrasco B.E., Frenk C.S., 1997,
	MNRAS, in press}
\myref{Glazebrook K., Peacock J.A., Collins C.A., Miller L., 1994,
	MNRAS, 266, 65}
\myref{Glazebrook K., Peacock J.A., Miller L., Collins C.A., 1995, 
	MNRAS, 275, 169}
\myref{Gioia I.M., Maccacaro T., Schild R.E., Wolter A., Stocke J.T.,
	Morris S.L. \& Henry J.P. 1990, ApJS, 72, 567}
\myref{Gioia I.M., Luppino G.A., 1994, ApJS, 94, 583}
\myref{Gunn J.E., Oke J.B., 1975, ApJ, 195, 255}
\myref{Haussman M.A., Ostriker J.P., 1978, ApJ, 224, 320}
\myref{Henry J.P., 1997, preprint}
\myref{Hoessel J.G., 1980, ApJ, 241, 493}
\myref{Hoessel J.G., Gunn J.E., Thuan T.X., 1980, ApJ, 241, 486}
\myref{Hoessel J.G., Schneider D.P., 1985, AJ, 90, 1648}
\myref{Huang J.-S., Cowie L.L., Gardner J.P., Hu E.M., Songaila A., 
	Wainscoat R.J., 1997, ApJ, in press}
\myref{Hubble E., 1929, Proc. Nat. Acad. Sci. 15, 168}
\myref{Hudson M.J., Ebeling H., 1997, ApJ, 479, 621} 
\myref{Humason M.L., Mayall N.U., Sandage A.R., 1956, AJ, 61, 97}
\myref{Kochanek C.S., 1996, ApJ, 466, 638}
\myref{Kristian J., Sandage A., Westphal J.A., 1978, ApJ, 221, 383}
\myref{Lauer T.R., Postman M., 1994, ApJ, 425, 418}
\myref{Lebofsky M.J., 1980, in Abell G.O., Peebles P.J.E., eds,
	Proc. IAU Symp. 92, Objects of High Redshift. Reidel,
	Dordrecht, p. 257}
\myref{Lebofsky M.J., Eisenhardt P.R.M., 1986, ApJ, 300, 151}
\myref{Lilly S.J., Longair M.S., 1984, MNRAS, 211, 833}
\myref{Mathis J.S., 1990, ARAA, 28, 37} 
\myref{McLeod B.A., Bernstein G.M., Rieke M.J., Tollestrup E.V., Fazio
	G.G., 1995, ApJS, 96, 117}
\myref{McNamara B.R., O'Connell R.W., 1992, ApJ, 393, 579}
\myref{Metcalfe N., Shanks T., Campos A., Fong R., Gardner J.P., 1996,
	Nat, 383, 236}
\myref{Miller G.E., Scalo J.M., 1979, ApJS, 41, 513}
\myref{Mushotsky R.F., Scharf C.A., 1997, ApJ, in press}
\myref{Mobasher B., Ellis R.S., Sharples R.M., 1986, MNRAS, 223, 11}
\myref{Peach J.V., 1970, ApJ, 159, 753}
\myref{Perlmutter S., et al., 1997, ApJ, 483, 565}
\myref{Postman M., Lauer T., 1995, ApJ, 440, 28}
\myref{Press W.H., Teukolsky S.A., Vetterling W.T., Flannery B.R., 1992,
	Numerical Recipies in FORTRAN, The Art of Scientific Computing, 
	Second Edition, Cambridge University Press, Cambridge}
\myref{Romer A.K., Collins C.A., B\"{o}hringer H., Cruddace R.G., Ebeling
	H., MacGillivray H.T., Voges W., 1994, Nature, 372, 75}
\myref{Salpeter E.E., 1955, ApJ, 121, 161}
\myref{Scalo J.M., 1986, FCPh, 11,1}
\myref{Sandage A.R., 1972a, ApJ, 173, 485}
\myref{Sandage A.R., 1972b, ApJ, 178, 1}
\myref{Sandage A.R., 1988, ARA\&A, 26, 561}
\myref{Sandage A.R., 1995, in Binggeli B., Buser R., eds, The Deep
	Universe, Saas-Fee Advanced Course 23. Springer, Berlin, p. 3.}
\myref{Sandage A.R., Hardy E., 1973, ApJ, 183, 743}
\myref{Sandage A.R., Kristian J., Westphal J.A., 1976, ApJ, 205, 688}
\myref{Stark A.A., Gammie C.F., Wilson R.W., Bally J., Linke R.A., 
	Heiles C., Hurwitz M., 1992, ApJS, 79,77}
\myref{Stocke J.T., Morris S.L., Gioia I.M., Maccacaro T., Schild R.E.,
	Wolter A., Fleming T.A., Henry J.P. 1991, ApJS, 76, 813}
\myref{Tr\"{u}mper J., 1993, Science, 260, 1769}
\myref{van Haarlem M.P., Frenk C.S., White S.D.M., 1997, MNRAS, 287, 817}
\myref{Voges W., 1992, Proceedings of Satellite Symposium 3, ESA ISY-3, p9}
\myref{Yates M.G., Miller L., Peacock J.A., 1986, MNRAS, 221, 311}

\appendix
\setcounter{table}{0}
\renewcommand\thesection{\Alph{section}}
\renewcommand\thetable{\thesection\arabic{table}}

\section{PARTIAL SPEARMAN TEST RESULTS}

In this Appendix we tabulate results from the partial
Spearman rank correlation analysis of Section 5 that it would be
too tedious to include in the main body of the text of this paper.
The partial Spearman rank correlation
coefficient $r_{\rm AB,C}$ (defined in equation 4 of Section 6) 
quantifies the strength of the correlation between $A$ and $B$, once
the effect of their individual correlations with $C$ has been removed.
The significance of $r_{\rm AB,C}$ may readily be computed. This is
the probability, $P_{\rm AB,C}$, that a correlation coefficient with
absolute value as high as $|r_{\rm AB,C}|$ would arise if the
correlation between $A$ and $B$ arose solely from their separate
correlations with $C$: in common with Yates et al. (1986), we find
that Monte Carlo simulations indicate that this may be computed
sufficiently accurately using the approximnation that 
\mbox{$r_{\rm AB,C} [(N-3)/(1-r_{\rm AB,C}^2)]^{1/2}$} is distributed
as Student's t statistic.

We tabulate here the $r_{\rm AB,C}$ values for the five
quantities studied in Section 5: (i) redshift, $z$; (ii) absolute
K band magnitude, $M_{\rm K}$, computed using the K-correction model
of equation (1); (iii) X-ray luminosity, $L_{\rm X}$; 
(iv) corrected near neighbour number, $\hat{N}$; and (v) BCG structure
parameter, $\alpha$. In each table, the upper half lists the partial 
Spearman rank correlation coefficient, $r_{\rm AB,C}$, while the lower
half gives \mbox{$\hat{P}_{\rm AB,C}\equiv-\log_{\rm 10}[P(r_{\rm
AB,C})]$}. We consider the effect of setting $C$ equal to
each of these in turn, by which we study the significance of
correlations between pairs of variables $A$ and $B$ after removing
the effect of correlations between $A$ and $C$ and between $B$ and
$C$.

\begin{table}

\caption{Results of partial Spearman rank correlation analysis for 
constant redshift, i.e. $C=z$. The upper
half of the table gives the values of
$r_{\rm AB,C}$, while the lower half gives \mbox{$\hat{P}_{\rm
AB} \equiv-\log_{\rm 10}[P(r_{\rm
AB,C})]$}, where $P(r_{\rm AB,C})$ is the probability that the absolute
value of the correlation coefficient would be as large as $|r_{\rm
AB,C}|$ for a sample of the same size, under the hypothesis that the
correlations between $A$ and $B$ arise solely from their individual
correlations with $C$}

\begin{tabular}{lcccccc}
 & $z$ & $M_{\rm K}$ & $L_{\rm X}$ & $\hat{N}$ & $\alpha$\\ 
$z$    & - & - & - & - & - \\
 $M_{\rm K}$   & - & - & -0.46 &-0.39  &-0.39  \\
 $L_{\rm X}$  & - &2.86  & - &0.42  & 0.32 \\
 $\hat{N}$  & - & 2.17 &2.45  & - &0.58  \\
 $\alpha$  & - &2.15  &1.54  &4.67  & - \\

\end{tabular}

\end{table}  

\begin{table} 

\caption{Results of partial Spearman rank correlation analysis for 
constant absolute magnitude, i.e. $C=M_{\rm K}$: format as for Table
A1.}

\begin{tabular}{lcccccc}
 & $z$ & $M_{\rm K}$ & $L_{\rm X}$ & $\hat{N}$ & $\alpha$\\ 
$z$    & - & - & 0.60 & -0.76 &0.01  \\
 $M_{\rm K}$   &  & - & - & - & - \\
 $L_{\rm X}$  & 4.91 & - & - & 0.19 & 0.15 \\
 $\hat{N}$  & 0.21 & - & 0.69 & - & 0.50\\
 $\alpha$  & 0.02 & - & 0.48 & 3.45 & - \\

\end{tabular}

\end{table}  

\begin{table} 

\caption{Results of partial Spearman rank correlation analysis for 
constant X-ray luminosity, i.e. $C=L_{\rm X}$: format as for Table
A1.}

\begin{tabular}{lcccccc}
 & $z$ & $M_{\rm K}$ & $L_{\rm X}$ & $\hat{N}$ & $\alpha$\\ 
$z$    & - & 0.12 & - & -0.26 &-0.13  \\
 $M_{\rm K}$   & 0.35 & - & - & -0.27 & -0.30 \\
 $L_{\rm X}$  & - & - & - & - & - \\
 $\hat{N}$  & 1.10 & 1.15 & - & - & 0.53 \\
 $\alpha$  & 0.40 & 1.37 & - & 3.84 & - \\

\end{tabular}

\end{table}  

\begin{table} 

\caption{Results of partial Spearman rank correlation analysis for 
constant corrected near neighbour number, i.e. $C=\hat{N}$:
format as for Table A1.}

\begin{tabular}{lcccccc}
 & $z$ & $M_{\rm K}$ & $L_{\rm X}$ & $\hat{N}$ & $\alpha$\\ 
$z$    & - & -0.23 & 0.65 & - & 0.11 \\
 $M_{\rm K}$   & 0.92 & - & -0.41 & - & -0.23 \\
 $L_{\rm X}$  & 5.97 & 2.33 & - & - & 0.15 \\
 $\hat{N}$  & - & - & - & - & - \\
 $\alpha$  & 0.31 & 0.94 & 0.49 & - & - \\

\end{tabular}

\end{table}  

\begin{table} 

\caption{Results of partial Spearman rank correlation analysis for 
constant structure parameter, i.e. $C=\alpha$: format as for
Table A1.}

\begin{tabular}{lcccccc}
 & $z$ & $M_{\rm K}$ & $L_{\rm X}$ & $\hat{N}$ & $\alpha$\\ 
$z$    & - & -0.20 & 0.62 & -0.04 & - \\
 $M_{\rm K}$   & 0.74 & - & -0.42 & -0.21 & - \\
 $L_{\rm X}$  & 5.34 & 2.40 & - & 0.21 & - \\
 $\hat{N}$  & 0.12 & 0.78 & 0.79 & - &- \\
 $\alpha$  & - & - & - & - & - \\

\end{tabular}

\end{table}  

\end{document}